\documentclass[journal,10pt]{IEEEtran}
\usepackage{amsfonts,amssymb}
\usepackage{amsmath}
\usepackage{array}
\usepackage{cite}
\usepackage{url}
\usepackage{xcolor}
\usepackage{cite,graphicx,amsmath,amssymb}
\usepackage{subfigure}
\usepackage{fancyhdr}
\usepackage{mdwmath}
\usepackage{mdwtab}
\usepackage{amsthm}
\usepackage{booktabs}
\usepackage{multirow}
\usepackage{longtable}
\usepackage{rotating}

\newtheorem{remark}{Remark}
\newtheorem{theorem}{Theorem}
\newtheorem{definition}{Definition}

\newtheorem{lemma}{Lemma}

\newtheorem{corollary}{Corollary}

\newtheorem{proposition}{Proposition}

\usepackage[breaklinks,colorlinks,linkcolor=black,citecolor=black,urlcolor=black]{hyperref}

\begin{document}

\title{STAR-RIS Aided NOMA in Multi-Cell Networks: A General Analytical Framework with Gamma Distributed Channel Modeling}

\author{Ziyi~Xie,~\IEEEmembership{Student Member,~IEEE,} Wenqiang~Yi,~\IEEEmembership{Member,~IEEE,} Xuanli~Wu,~\IEEEmembership{Member,~IEEE,} Yuanwei~Liu,~\IEEEmembership{Senior Member,~IEEE,} and Arumugam~Nallanathan,~\IEEEmembership{Fellow,~IEEE}

%\thanks{This work was supported in part by the Engineering and Physical Sciences Research Council (EPSRC), U.K., under Grant EP/R006466/1, in part by the National Natural Science Foundation of China under Grant 61971161, in part by Heilongjiang Touyan Team under Grant HITTY-20190009, and in part by the China Scholarship Council under Grant 202106120107. ({\it Corresponding author: Xuanli Wu.})}

\thanks{Z. Xie and X. Wu are with the School of Electronics and Information Engineering, Harbin Institute of Technology, Harbin 150001, China (email: \{ziyi.xie, xlwu2002\}@hit.edu.cn).}

\thanks{W. Yi, Y. Liu, and A. Nallanathan are with the School of Electronic Engineering and Computer Science, Queen Mary University of London, E1 4NS, U.K. (email: \{w.yi, yuanwei.liu, a.nallanathan\}@qmul.ac.uk).}
 
 }

\maketitle
\begin{abstract}
The simultaneously transmitting and reflecting reconfigurable intelligent surface (STAR-RIS) is capable of providing full-space coverage of smart radio environments. This work investigates STAR-RIS aided downlink non-orthogonal multiple access (NOMA) multi-cell networks, where the energy of incident signals at STAR-RISs is split into two portions for transmitting and reflecting. We first propose a fitting method to model the distribution of composite small-scale fading power as the tractable Gamma distribution. Then, a unified analytical framework based on stochastic geometry is provided to capture the random locations of RIS-RISs, base stations (BSs), and user equipments (UEs). Based on this framework, we derive the coverage probability and ergodic rate of both the typical UE and the connected UE. In particular, we obtain closed-form expressions of the coverage probability in interference-limited scenarios. We also deduce theoretical expressions in conventional RIS aided networks for comparison. The analytical results show that optimal energy splitting coefficients of STAR-RISs exist to simultaneously maximize the system coverage and ergodic rate. The numerical results demonstrate that: 1) STAR-RISs are able to meet different demands of UEs located on different sides; 2) STAR-RISs with appropriate energy splitting coefficients outperform conventional RISs in the coverage and the rate performance.
\end{abstract}


\section{Introduction}
Requirements for high data rates and heterogeneous services in future sixth-generation (6G) wireless networks bring challenges to system designs \cite{06G1,0EE,0surveyMarco}. The smart radio environment (SRE) is envisioned to be a promising solution \cite{0smartZ,0surveyMarco}. Equipped with several low-cost reconfigurable elements and a smart controller, a reconfigurable intelligent surface (RIS) is capable of intelligently altering the phase of signals \cite{0surveyMarco,0surveyLiu}, hence the propagation of which is controllable and the SRE can be realized. However, the main issue of conventional reflecting-only RISs in existing works is that user equipment (UE) can only receive reflected signals from base stations (BSs) located on the same side of the assisted RIS, which degrades the coverage performance, especially for those blocked UEs. Thanks to the recent development of metasurfaces, the concept of simultaneous transmitting and reflecting RISs (STAR-RISs) has been proposed, where incident signals can not only be reflected within the same half-space in front of the RIS but can be refracted to the other half-space \cite{0STAR360,0STARRISMu,0omni1}. Thus, STAR-RISs are able to provide full-space coverage of SRE. 

As stated in \cite{0STAR360}, there are three practical operating protocols for STAR-RISs, namely energy splitting, mode switching, and time switching. In energy splitting and mode switching protocols, since the incident signal is split into two portions by the STAR-RIS, a multiple access scheme is required to distinguish these two parts for successful demodulation at UEs located on different sides of the STAR-RIS. Compared with orthogonal multiple access (OMA), non-orthogonal multiple access (NOMA) has been considered to be a competent technique due to its ability for spectral efficiency enhancement and UE fairness guarantee \cite{0NOMA1,0NOMA3}. The key idea of the NOMA scheme is to serve multiple UEs in the same resource block (RB) by employing superposition coding and successive interference cancellation (SIC) at transmitters and receivers, respectively. On the other hand, the deployment of STAR-RISs is beneficial to NOMA systems. For NOMA UEs with weak channel conditions, STAR-RISs are able to create stronger transmission links. Moreover, since STAR-RISs have the capability of adjusting channel gains of different NOMA UEs, they can offer flexible decoding orders according to the priority of UEs.
\subsection{Related Works}
For RIS aided networks, initial research contributions have paid attention to the performance analysis in single-cell systems. In these works, the channel modeling for RIS assisted communications is firstly investigated as it plays an important role when theoretically evaluating the enhancements and limitations of RISs. The authors in \cite{1Erik} derived the far-field path loss expression based on physical optics techniques and pointed out that the path loss value is correlated to the product of two distances of the cascaded link. In \cite{1Cui}, the authors obtained free-space path loss in both near-field and far-field cases. Experimental measurements were also carried out to validate the accuracy of the analysis results. 
Considering the small-scale fading, most existing works utilized approximations to characterize the composite channel gain, where RISs are regarded as integrated antennas \cite{1CLT1,1CLT2,1convolution1,1convolution2}. The authors in \cite{1CLT1} and \cite{1CLT2} assumed that the number of RIS elements is sufficiently large, and hence the central limit theorem (CLT) was applied to approximate the distribution of the channel gain. After that, the system capacity and the spatial throughput were derived in \cite{1CLT1} and \cite{1CLT2}, respectively. For an arbitrary number of elements, the authors in \cite{1convolution1} and \cite{1convolution2} employed the convolution theorem to evaluate the asymptotic outage probability in STAR-RIS aided networks. A curve fitting method was also proposed in \cite{1convolution2}. Different from the above works, the authors in \cite{1exact} derived the exact coverage probability using Gil-Pelaez inversion, where Nakagami-$m$ fading was assumed.

Recently, the system performance of RIS aided multi-cell networks has been evaluated. In \cite{1multiXYG}, the authors simultaneously optimized the coverage and capacity in a two-cell system. The authors in \cite{1multimaxmin} considered a multi-cell multiple-input single-output network, where transmit and reflective beamforming vectors were jointly optimized to maximize the minimum weighted signal-to-interference-plus-noise ratio (SINR) at UEs. For large-scale deployment scenarios, system optimizations were investigated in \cite{1multiopt1} and \cite{1multiopt2}. In \cite{1multiopt1}, the optimal association solution among BSs, RISs, and UEs was obtained for maximizing the utility of the considered system. In \cite{1multiopt2}, the authors focused on the capacity improvement in a cell-free structure. These works optimized the system parameters in particular setups with fixed BSs and RISs. To characterize the randomness property of large-scale networks, stochastic geometry is an efficient tool \cite{1WYi}, which has been widely utilized to evaluate the average performance of multi-cell networks with largely deployed RISs \cite{1gammaappr,1multicell1,1multicell2}. However, in this scenario, channel models proposed in a single-cell setup have to be further simplified to tractable formats. A recent work \cite{1gammaappr} considered double-Rayleigh fading and approximated the composite channel gain as the Gamma distribution. Besides, based on a tractable linear RIS model proposed in \cite{1rislinear}, the authors in \cite{1multicell1} and \cite{1multicell2} analyzed the coverage probability and rate performance.

Motivated by the benefits including high spectral efficiency and the flexible SIC order from the integration of RISs and NOMA, recent research efforts have been devoted to RIS-enabled NOMA systems. System optimizations were considered in \cite{1optcov,1optsumrate,1optenergy1,1optenergy2}. The authors in \cite{1optcov} maximized the area of the cell coverage by optimizing RIS placement. In \cite{1optsumrate}, the authors proposed a joint design to maximize the achievable system sum rate. Multiple parameters including beamforming vectors and power allocation coefficients were jointly optimized for the total transmit power minimization in \cite{1optenergy1} and \cite{1optenergy2}. By leveraging stochastic geometry, the spatial effects of large-scale RIS deployment were evaluated in both single-cell networks \cite{1NOMAsingle} and multi-cell networks \cite{1multicell1,1multicell2}. Additionally, the authors in \cite{1comp} investigated the performance enhancement of coordinated multipoint transmissions in a two-cell setup. However, all these works adopted conventional reflecting-only RISs, and the research on STAR-RIS aided NOMA networks is scarce. In STAR-RIS enhanced NOMA transmissions, optimization problems focused on sum rate maximization \cite{1optSTARrate} and optimality gap minimization \cite{1optSTARgap} were considered. For theoretical analysis, a recent work \cite{1convolution2} first evaluated three STAR-RIS operating protocols in a NOMA single-cell network. 

\subsection{Motivations and Contributions}
As we have discussed previously, NOMA schemes are able to enhance the spectral efficiency for STAR-RISs aided networks, and STAR-RISs have the potential to offer full-space coverage as well as decoding flexibility for NOMA systems. Although some initial works have validated the enhancement of STAR-RISs, most of them focused on specific small-scale fading environments, and the theoretical performance in large-scale deployment scenarios has not been investigated yet. One of the main difficulties is to characterize the composite RIS aided channel as a tractable expression. In the prior work \cite{1convolution2}, the authors utilized the curve fitting tool in Matlab to fit the STAR-RIS aided Rician fading channel under a single-cell setup.
In this work, we first theoretically provide a tractable and accurate expression to characterize the composite channel model with general small-scale fading. To shed light on the performance improvement brought by STAR-RISs in multi-cell networks, a stochastic geometry-based analytical framework for a general case is then developed. The main contributions are summarized as follows:
\begin{itemize}
	\item We derive a general expression to characterize the distribution of the small-scale fading power of the composite channel including multiple independent RIS-based channels. By exploiting the CLT and the method of moment, the channel power gain (CPG) of the RIS aided link can be approximated by the Gamma distribution, whose parameters are only related to the mean value and the variance of the considered small-scale fading model. 
    Due to the channel hardening effect, the asymptotic value of CPG for the desired signal is a constant when $N \to \infty$.
	\item Considering downlink transmissions, we develop an analytical framework for the STAR-RIS aided NOMA multi-cell networks based on stochastic geometry, where the distributions of BSs, STAR-RISs, and UEs are independent homogeneous Poisson point processes (PPPs). In this framework, STAR-RISs are employed to assist the blocked typical UE and connected UE in the NOMA UE pair to communicate with their BS. By limiting the locations of BSs within the same half-space of the typical UE, this framework can be applied to conventional RIS aided networks.
	\item Focusing on the energy splitting protocol, we evaluate the coverage performance and ergodic rate for this STAR-RIS aided network. Using a novel analytical method, we derive the theoretical expressions of these two metrics for both the typical UE and the connected UE. In particular, the interference-limited case is considered as a special case, where we obtain closed-form expressions for the coverage probability. We also provide expressions in conventional RIS aided networks for comparison. Besides, the impact of the energy splitting coefficients is investigated. The analytical results demonstrate that the system performance can be improved by adjusting the energy splitting coefficients. 
	\item The numerical results validate our theoretical analysis and illustrate that: 1) the NOMA scheme significantly enhances the ergodic sum rate as well as the coverage performance for the connected UE in NOMA systems; 2)  STAR-RISs with appropriate energy splitting coefficients outperform conventional RISs in both coverage and rate performance; 3) STAR-RISs bring flexibility to NOMA systems by reconfigurable energy splitting coefficients.
\end{itemize}

\subsection{Organizations}
The rest of this paper is organized as follows. In Section II, we introduce the system model of the STAR-RIS aided NOMA multi-cell networks that we consider. 
In Section III, we provide a fitting method to characterize the small-scale fading for general cases. In section IV, we derive the analytical expressions of the coverage probability. In Section V, we derive the analytical expressions of the ergodic rate. 
Section VI presents numerical results. Finally, we draw the conclusions in Section VII.

\section{System Model}
\subsection{Network Model}
\begin{figure*} [t!]
	\centering
	\includegraphics[width = 4.3 in] {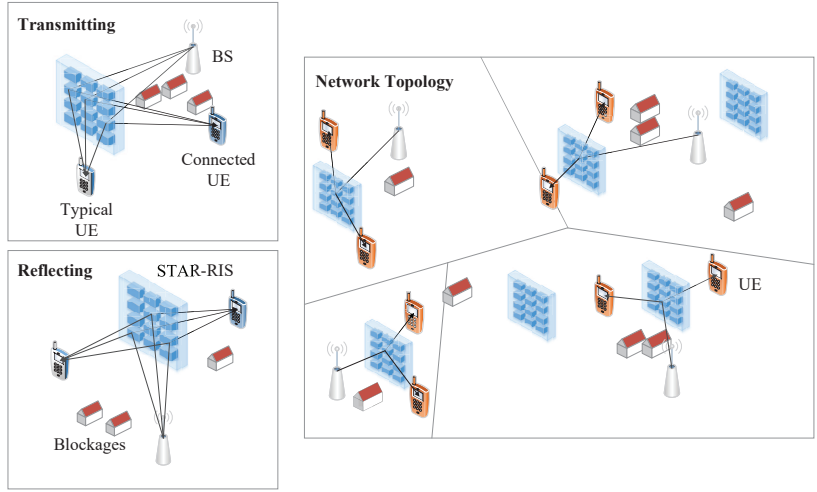}
	\caption{Illustration of the considered STAR-RIS aided networks: the left two subfigures show the assistant STAR-RIS transmitting and reflecting signals for the typical UE, respectively; the right subfigure depicts the network topology of the STAR-RIS aided networks.}
	\label{fig: system model}
\end{figure*}
In this paper, STAR-RIS aided downlink NOMA multi-cell networks are considered.
The locations of BSs, STAR-RISs, and UEs obey three independent homogeneous PPPs $\Phi_B$, $\Phi_R$, and $\Phi_U$ in $\mathbb{R}^2$ with density $\lambda_B$, $\lambda_R$, and $\lambda_U$. Both BSs and UEs are equipped with a single antenna. The transmit power of BSs is $P_B$. The STAR-RIS consists of $N$ reflecting elements, all of which are able to simultaneously transmit and reflect signals. The energy splitting protocol is considered in this work.

To improve the spectral efficiency, two NOMA UEs are grouped in each orthogonal RB. We define a UE randomly selected from $\Phi_U$ as the typical UE $u_t$ and set the location of the typical UE as the origin of the considered plane. The typical UE associates to its BS with the aid of a STAR-RIS. The other one of the paired UEs $u_c$, called the connected UE, is located at the opposite side of the STAR-RIS to the typical UE, jointing the same RB of the typical UE to form the typical NOMA pair. Therefore, the desired signal is split by the STAR-RIS, being transmitted and reflected to the typical paired UEs, respectively. For tractable analytical expressions, simple models are employed, which obey the following assumptions:

{\it Assumption 1:} All direct links between BSs and the typical UE are blocked \footnote{This assumption is because the impact of the direct BS-UE link can be ignored when the number of STAR-RIS elements is large \cite{2dirpath}.}. Therefore, the communication between the serving BS and the typical UE is assisted by STAR-RISs.

{\it Assumption 2:} The connected UE is the target UE and it has been associated with the BS through the assisted STAR-RIS in the previous UE association process. Therefore, the distance between the connected UE and the STAR-RIS $d_c$ is known at the serving BS. The typical UE is an add-on UE, which is served by the same RB via NOMA.

{\it Assumption 3:} The QoS-based SIC \cite{2assumpt} is considered in the NOMA scheme. To guarantee the communication quality of the connected UE, more power is allocated to the connected UE, and the SIC is always processed at the typical UE.

{\it Assumption 4:} Ideal STAR-RISs are considered, which are capable of independently controlling the transmitted and reflected signals.

\subsection{Channel Model}
Since all direct BS-UE links are blocked for the typical UE, the desired signal is transmitted/reflected by an assisted STAR-RIS. Detailed descriptions of the STAR-RIS aided link are provided in the following.

We use $u_\epsilon$ to denote any UE in the typical NOMA pair, where the subscript $\epsilon \in \{t,c\}$ presents the type of UEs. Since STAR-RISs are regarded as integrated antennas in this work, there are $N$ communication channels between the BS $i$ and the UE $u_\epsilon$. According to the mechanism of the energy splitting protocol, the incident signal at each STAR-RIS element is split into two parts for transmitting and reflecting, respectively. We denote ${\bf{\Theta }}_{\chi} = \sqrt {\beta _\chi } {\rm diag} \left( {{e^{j\theta _{\chi,1} }},{e^{j\theta _{\chi,2} }},...,{e^{j\theta _{\chi,N} }}} \right)$ as the transmissive/reflctive-coefficient matrix of the STAR-RIS, where the subscript $\chi$ denote the transmission mode for the signal, i.e., $\chi = {\rm T}$ and $\chi = {\rm R}$ represent transmitting and reflecting signals, respectively. In particular, $j=\sqrt{-1}$, $\theta_{\chi,n} \in [0,2\pi)$ with $n \in  \{1,2,...,N\}$, and $\beta _\chi \in [0,1]$ is the energy splitting coefficient. As STAR-RISs are passive and other energy consumptions are assumed to be negligible, we have $\beta _{\rm T} + \beta_{\rm R} = 1$. In this work, $\beta_\chi$ on all elements are assumed to be the same. This setting only needs low-complexity hardware and the DOCOMO’s smart glass model is such a STAR-RIS prototype. In practice, $\beta_\chi$ of this prototype can be tuned by adjusting the distance between substrates \cite{2controlbeta}. We also denote the small-scale fading vectors of the BS-RIS link and the RIS-UE link as ${\bf{H}}_{ BR} = [h_{BR,1},h_{BR,2},...,h_{BR,N}] \in \mathbb{C}^{N \times 1}$ and ${\bf{H}}_{RU} = [h_{RU,1},h_{RU,2},...,h_{RU,N}] \in \mathbb{C}^{N \times 1}$, respectively. The overall channel gain from the BS $i$ to the UE $u_\epsilon$ assisted by the STAR-RIS $k \in \Phi_R$ can be expressed as
\begin{align}
{H}_{\epsilon,i}^{(k)} = \sqrt{C_r\left(r_{i}^{(k)} d_{\epsilon}^{(k)} \right)^{-\alpha _r}} \left( {\bf{H}}_{RU} \right)^{\rm H} {\bf{\Theta }}_{\chi} {\bf{H}}_{BR},
\end{align}
where $L_{\epsilon,i}^{(k)} = C_r\left(r_{i}^{(k)} d_{\epsilon}^{(k)} \right)^{-\alpha _r}$ is the path loss of the STAR-RIS aided link. The $C _r$ is the intercept. The $\alpha_r$ is the path loss exponent. $r_{i}^{(k)}$ denotes the distance between the BS and the assisted STAR-RIS. $d_{\epsilon}^{(k)}$ represents the distance between the STAR-RIS and the UE $u_\epsilon$.

\subsection{UE Association and Channel Power Gain Characterization}\label{section: association model}
For the typical UE $u_t$, the closest association criterion \cite{2closeassociate} is employed. Specifically, the typical UE associates to its nearest STAR-RIS, and the STAR-RIS chooses the nearest BS as the serving BS. The probability density function (PDF) of the serving distance can be given by
\begin{align}\label{eq: RU pdf}
	f_{RU}(x) = 2\pi\lambda_Rx \exp(-\pi\lambda_Rx^2),
\end{align}
\begin{align}\label{eq: BR pdf}
	f_{BR}(x) = 2\pi\lambda_Bx \exp(-\pi\lambda_Bx^2).
\end{align}

Let ${h}_{r}$ denote the equivalent overall small-scale fading for the STAR-RIS aided composite channel of the UE, whose power is given by
\begin{align}
|{h}_{r}|^2 \triangleq \left| \left( {\bf{H}}_{RU} \right)^{\rm H} {\tilde{{\bf{\Theta }}}} _{\chi} {\bf{H}}_{BR} \right|^2 ,
\end{align}
where ${\tilde{{\bf{\Theta }}}} _{\chi} = {\bf{\Theta }}_{\chi} / \sqrt{ \beta _{\chi}} = {\rm diag} \left( {{e^{j\theta _{\chi,1} }},{e^{j\theta _{\chi,2} }},...,{e^{j\theta _{\chi,N} }}} \right)$ is the normalized phase-shifting matrix of the STAR-RIS. According to \cite{1CLT2}, the channel phase $\angle \left(h_{BR,n} h_{RU,n} \right)$ can be obtained from the channel estimation. 
To achieve the maximum received power at the receiver, the STAR-RIS reconfigures the phase shifts $\theta_{\chi,n}=-\angle \left(h_{BR,n} h_{RU,n} \right)$ so that signals from all channels are of the same phase at the UE. As a result, the power gain of the small-scale fading for the signal is
	\begin{align}
		|{h}_{r,S}|^2 = \left( \left| \left( {\bf{H}}_{RU} \right)\right|^{\rm H}  \left|{\bf{H}}_{BR} \right| \right)^2 = \left(\sum\limits_{n = 1}^N \left| h_{BR,n}\right|\cdot\left| h_{RU,n} \right| \right)^2.
\end{align}

We denote the equivalent overall small-scale fading of the scatter interference as $h_{r,I}$, whose power is given by
 \begin{align}
 	|{h}_{r,I}|^2 = \left(\sum\limits_{n = 1}^N \left| h_{BR,n}\right|\cdot\left| h_{RU,n} \right| e^{j\theta_{I,n}} \right)^2,
 \end{align}
where $\theta_{I,n} \in [0,2\pi)$ is the phase at the UE. Then the overall CPG for the typical UE can be expressed as $|{H}_{t,i,\kappa}^{(k)}|^2 = \beta _\chi L_{t,i}^{(k)}|h_{r,\kappa}|^2$ for $\kappa \in \{S,I\}$.

The connected UE is served by the same STAR-RIS as the typical UE but located on the other side, and the distance between the assisted STAR-RIS and the connected UE is fixed. Thus, $d_c^{(k)} = d_c$ is a constant. Similarly, the overall CPG for the connected UE is hence given by $|{H}_{c,i,\kappa}^{(k)}|^2 = \left(1-\beta _\chi \right)L_{c,i}^{(k)}|h_{r,\kappa}|^2$ for $\kappa \in \{S,I\}$.

\subsection{SINR Analysis}
Considering power domain NOMA, let $a_t$ and $a_c$ denote the power allocation coefficients for the typical UE and the connected UE, respectively. Thus, we have $a_t<a_c$ and $a_t+a_c = 1$. 

The typical UE first decodes the information of the connected UE in the typical NOMA group with the following SINR
\begin{align}
\gamma _{t \to c} = \frac{a_c P_B \beta _\chi L_{t,i}^{(k)} |h_{r,S}|^2} {a_t P_B \beta _\chi L_{t,i}^{(k)} |h_{r,S}|^2 + I_{t} + {n_0} ^2},
\end{align}
with
\begin{align}
I_t = \sum\limits_{m \in \Phi_B^{\rm T} \backslash i}  P_B \beta _{\rm T} L_{t,m}^{(k)} |h_{r,I}|^2 + \sum\limits_{m \in \Phi_B^{\rm R} \backslash i}  P_B \beta _ {\rm R} L_{t,m}^{(k)} |h_{r,I}|^2,
\end{align}
where ${n_0} ^2$ is the additive white Gaussian noise (AWGN) power and $I_t$ is the interference from the serving STAR-RIS $k$.
The $I_t$ consists of two portions: interference transmitted and reflected by the assisted STAR-RIS $k$. We use $\Phi_B^{\rm R}$ to denote BSs located on the same side of the STAR-RIS $k$ as the typical UE, and hence the typical UE only receives reflected signals from these BSs. Similarly, we use $\Phi_B^{\rm T}$ to represent the BS sets of transmitting. For tractability, we only consider the impact of STAR-RIS $k$ and ignore the interference from the other STAR-RISs. Therefore, the performance obtained in this work can be regarded as an upper bound.

After the SIC process, the decoding SINR at the typical UE can be expressed as
\begin{align}\label{eq: SINR typical}
\gamma _{t} = \frac{a_t P_B \beta _\chi L_{t,i}^{(k)} |h_{r,S}|^2} {I_{t} + {n_0} ^2}.
\end{align}

For the connected UE, the signal can be decoded by treating the message transmitted to the typical UE as interference. Therefore, the decoding SINR at the connected UE is as follows
\begin{align}\label{eq: SINR connected}
\gamma _{c} = \frac{a_c P_B (1-\beta_{\chi}) L_{c,i} ^{(k)} |h_{r,S}|^2} {a_t P_B (1-\beta_{\chi}) L_{c,i}^{(k)} |h_{r,S}|^2 + I_{c} + {n_0} ^2},
\end{align}
with
\begin{align}
I_c = \sum\limits_{m \in \Phi_B^{\rm T} \backslash i}  P_B \beta _{\rm R} L_{c,m}^{(k)} |h_{r,I}|^2 + \sum\limits_{m \in \Phi_B^{\rm R} \backslash i}  P_B \beta _ {\rm T} L_{c,m}^{(k)} |h_{r,I}|^2,
\end{align}
where $I_c$ is the interference for the connected UE.

\section{Fitting the Composite Small-Scale Fading Power}\label{Sec: Channel}
For the STAR-RIS aided link, multiple elements of the STAR-RIS introduce the composite channel, the accurate power of which is intractable for performance analysis in large-scale deployment multi-cell scenarios. In this section, we first provide a tractable fitting method to characterize the distribution of the composite CPG for a general fading case. Some typical cases are then investigated, and the fitting results are validated at last.

\subsection{General Small-Scale Fading Model}
We begin by considering a general expression of $h_{r,n} = \left|h_{BR,n}\right|\cdot\left| h_{RU,n}\right|$. We use $\mu _r$ and ${\sigma _r}^2$ to denote the mean and variance of $h_{r,n}$, respectively. Then we can provide the approximated distribution of the composite small-scale fading CPG for the desired signal as follows.
\begin{lemma}\label{lemma: Gamma fitting}
For the desired signal, the distribution of the overall small-scale fading CPG of the STAR-RIS aided link can be approximated by a Gamma distribution
\begin{align}
	|{h}_{r,S}|^2 \sim {\Gamma} \left( \frac{{M_r}^2}{V_r}, \frac{V_r}{M_r} \right),
\end{align}
where $M_r = {\mu _r}^2N^2 +  {\sigma _r}^2N$ and $V_r = 4{\mu _r}^2{\sigma _r}^2N^3 + 2{\sigma _r}^4N^2$.
\end{lemma}

\begin{IEEEproof}
Noticed that the small-scale fading for $N$ different channels is independently and identically distributed, the CLT can be employed. Since signal phases from $N$ channels are aligned, the distribution of the composite channel gain obeys Gaussian distribution $\left|{h}_{r,S}\right| \sim {\cal N} \left( N\mu _r, N{\sigma _r} ^2 \right)$.

For simplicity, we denote $\mu_N =  N\mu _r$ and ${\sigma _N} ^2 =  N{\sigma _r} ^2$.
Thus, the power of this equivalent small-scale fading $|{h}_{r,S}|^2$ obeys noncentral chi-square distribution with the mean $\mathbb{E}[|{h}_{r,S}|^2] = {\mu_N}^2 +  {\sigma _N} ^2$. Considering the fourth order moment of $h_{r,S}$ is $\mathbb{E}[|{h}_{r,S}|^4] = {\mu _N}^4 + 6{\mu _N}^2{\sigma _N}^2 + 3{\sigma _N}^4$, the variance can be calculated by ${\rm var}[|{h}_{r,S}|^2] = \mathbb{E}[|{h}_{r,S}|^4] - (\mathbb{E}[|{h}_{r,S}|^2])^2$. Using the method of moments, the distribution of $|h_{r,S}|^2$ can be approximated by a Gamma distribution $\Gamma (k_r,\theta_r)$ with the shape parameter and the scale parameter expressed as $k_r = (\mathbb{E}[|{h}_{r,S}|^2])^2 /{{\rm var}[|{h}_{r,S}|^2]}$ and $\theta_r = {{\rm var}[|{h}_{r,S}|^2]}/\mathbb{E}[|{h}_{r,S}|^2]$, respectively.
After some algebraic manipulations, this lemma is proved.
\end{IEEEproof}

When the number of STAR-RIS elements is large, we can obtain the following corollary.
\begin{corollary}\label{corollary: gamma fitting large N}
When $N$ is sufficiently large, the Gamma distribution in {\bf \textbf{Lemma \ref{lemma: Gamma fitting}}} is rewritten as
\begin{align}\label{eq: gamma fitting large N}
	|h_{r,S}|^2 \sim \Gamma \left( \frac{{\mu _r}^2}{4{\sigma _r}^2}N, 4{\sigma _r}^2N \right).
\end{align}
\end{corollary}

\begin{IEEEproof}
We can calculate that $\frac{{M_r}^2}{V_r} = \frac{{\mu _r}^2}{4{\sigma _r}^2}N + o(1)$ and $\frac{V_r}{M_r} = 4{\sigma _r}^2N + o(1)$. Then \eqref{eq: gamma fitting large N} is obtained.
\end{IEEEproof}

\begin{remark}
It can be found from \emph{\textbf{Corollary \ref{corollary: gamma fitting large N}}} that both the shape and scale parameters are in proportion to the number of STAR-RIS elements. Furthermore, let us recall
the property of the Gamma distribution, by which $\Gamma \left( \frac{{\mu _r}^2}{4{\sigma _r}^2}N, 4{\sigma _r}^2N \right) = {{\mu_r}^2N^2}  \Gamma \left( \frac{{\mu _r}^2N}{4{\sigma _r}^2}, \frac{4{\sigma _r}^2}{{\mu _r}^2N} \right)$. As $N \to \infty$, we have
\begin{align}
	\frac{|h_{r,S}|^2}{\mathbb{E}[|h_{r,S}|^2]} \sim \Gamma \left( \frac{{\mu _r}^2N}{4{\sigma _r}^2}, \frac{4{\sigma _r}^2}{{\mu _r}^2N} \right) \to 1 ,
\end{align}
which shows the channel hardening effect of the STAR-RIS aided link. Therefore, with the increase of $N$, the CPG asymptotically approaches a deterministic value.
\end{remark}

Similary, the CPG for the interference is obtained as follows.
\begin{lemma}\label{lemma: Gamma fitting interf}
For the interference signal, the distribution of the overall small-scale fading CPG of the STAR-RIS aided link is given by
\begin{align}
	|{h}_{r,I}|^2 \sim {\Gamma} \left( 1, N({\mu_r}^2+{\sigma_r}^2) \right) \triangleq {\Gamma} \left( 1, N{\sigma_I}^2 \right).
\end{align}
\end{lemma}
\begin{IEEEproof}
Note that the phases from different channels are uniformly random in $[0, 2\pi)$, the distribution of the composite channel gain obeys the complex Gaussian distribution according to \cite[Proposition 2]{1gammaappr}, i.e., ${h}_{r,I} \sim {\cal CN} \left(0, N{\sigma_I}^2 \right) = \sqrt{N} {\sigma_I} {\cal CN} \left(0, 1 \right)$. Then we have $|{h}_{r,I}|^2 \sim N{\sigma_I}^2 \Gamma(1,1) $ and the proof is completed.
\end{IEEEproof}
\begin{remark}
According to {\bf Remark 1} and {\bf Lemma 2}, when $N$ is large, the expectations of the $|h_{r,S}|^2$ and $|h_{r,I}|^2$ have a positive corrrelation with $N^2$ and $N$, respectively, i.e., $\mathbb{E}[|h_{r,S}|^2] \sim o(N^2)$ and $\mathbb{E}[|h_{r,I}|^2] \sim o(N)$. In this case, the noise is ignorable and the received SINR at UEs has a linear correlation with the number of RIS elements, i.e., $\gamma_\epsilon \sim o(N)$ for $\epsilon \in \{t,c\}$. 
\end{remark}

\vspace{-0.5 cm}
\subsection{Case Studies}\label{section: case study}
\begin{table*}
	\centering
	\caption{Typical Small-Scale Fading Models} \label{table: channel models}	
	\begin{tabular}{
			|m{1.7cm}<{\centering} |m{2.5cm}<{\centering}
			|m{2.35cm}<{\centering} |m{2.8cm}<{\centering} 
			|m{4.8cm}<{\centering}|}
		\hline
		Models & Channel Parameters & $\mu _r$ & ${\sigma _r}^2$ & Gamma Distributions ($N$ is large)  \\ \hline
		Rayleigh Channel & $\delta_1 >0$ & $\delta_1 \sqrt{\frac{\pi}{2}}$ & $\frac{4-\pi}{2}{\delta_1} ^2$ & $k_r = \frac{\pi}{4(4-\pi)}N$, $\theta_r = 2(4-\pi){\delta_1}^2 N$ \\ \hline
		Nakagami-$m$ Channel & $m_2 \ge \frac{1}{2}$, $\Omega_2 > 0$ & $\frac{\Gamma(m_2 + \frac{1}{2})}{\Gamma (m_2)}\left( \frac{\Omega_2}{m_2} \right) ^{\frac{1}{2}} $ & $\Omega_2 - \frac{\Omega_2}{m_2} \left( \frac{\Gamma(m_2 + \frac{1}{2})}{\Gamma (m_2)} \right) ^2 $ & $k_r = \frac{\Gamma(m_2 + \frac{1}{2})^2 }{4 \left(m_2\Gamma(m_2)^2 - \Gamma(m_2 + \frac{1}{2})^2 \right)}N$, $\theta_r = 4\Omega_2 N - \frac{4\Omega_2}{m_2} \left( \frac{\Gamma(m_2 + \frac{1}{2})}{\Gamma (m_2)} \right) ^2 N$ \\ \hline
		Rician Channel & $K_3 >0$, $\delta_3 >0$, $c_3>0$ & $\delta_3 \sqrt{\frac{\pi}{2}} \sqrt{\frac{1}{K_3+1}} + c_3\sqrt{\frac{K_3}{K_3+1}}$ & $\frac{4-\pi}{2}\frac{{\delta_3} ^2}{K_3+1}$ & $k_r = \frac{\pi {\delta_3}^2 + 4\delta_3 c_3 \sqrt{\frac{\pi K_3}{2}} + 2K_3 {c_3}^2}{4 (4-\pi) {\delta_3} ^2}N$, $\theta_r = \frac{2(4-\pi){\delta_3} ^2}{K_3+1}N$ \\ \hline
		Weibull Channel & $k_4 >0$, $\lambda_4 >0$ & $\lambda_4 \Gamma (1+\frac{1}{k_4})$ & ${\lambda_4} ^2 \left( \Gamma (1+\frac{2}{k_4}) - \right.$ $\left. \Gamma (1+\frac{1}{k_4})^2 \right)$ & $k_r = \frac{\Gamma (1+ \frac{1}{k_4})^2}{\Gamma (1+\frac{2}{k_4}) - \Gamma (1+\frac{1}{k_4})^2} \frac{N}{4}$, $\theta_r = 4{\lambda_4} ^2 \left( \Gamma (1+\frac{2}{k_4}) - \Gamma (1+\frac{1}{k_4})^2 \right)N$ \\ \hline
		Double-Rayleigh Channel & $\delta_5 >0$, $\delta_6 >0$ & $\frac{\pi \delta_5 \delta_6}{2}$ & $4(1- \frac{\pi^2}{16}){\delta_5} ^2 {\delta_6} ^2$ & $k_r = \frac{\pi ^2}{64(1 - \pi^2 /16)}N$, $\theta_r = (16-\pi^2) {\delta_5} ^2 {\delta_6} ^2 N$ \\ \hline
		Double-Rician Channel & $K_7 >0$, $K_8 >0$, $\delta_7 >0$, $\delta_8 >0$, $c_7>0$, $c_8>0$ & $\frac{\pi\delta_7\delta_8}{2} \sqrt{\frac{1}{\hat{K}_7\hat{K}_8}} + c_8\delta_7 \sqrt{\frac{\pi K_8}{2 \hat{K}_7\hat{K}_8}} + c_7\delta_8 \sqrt{\frac{\pi K_7}{2 \hat{K}_7\hat{K}_8}} + c_7c_8 \sqrt{\frac{K_7K_8}{\hat{K}_7\hat{K}_8}}$  & $\mathbb{E} \left[ \left( h_{r,n} \right)^2  \right] -{\mu_r}^2$   & eq. \eqref{eq: gamma fitting large N} \\ \hline
	\end{tabular}
\end{table*}
\begin{figure*}[!t] 
	\centering
	\subfigure[]{\includegraphics[width=3.1in]{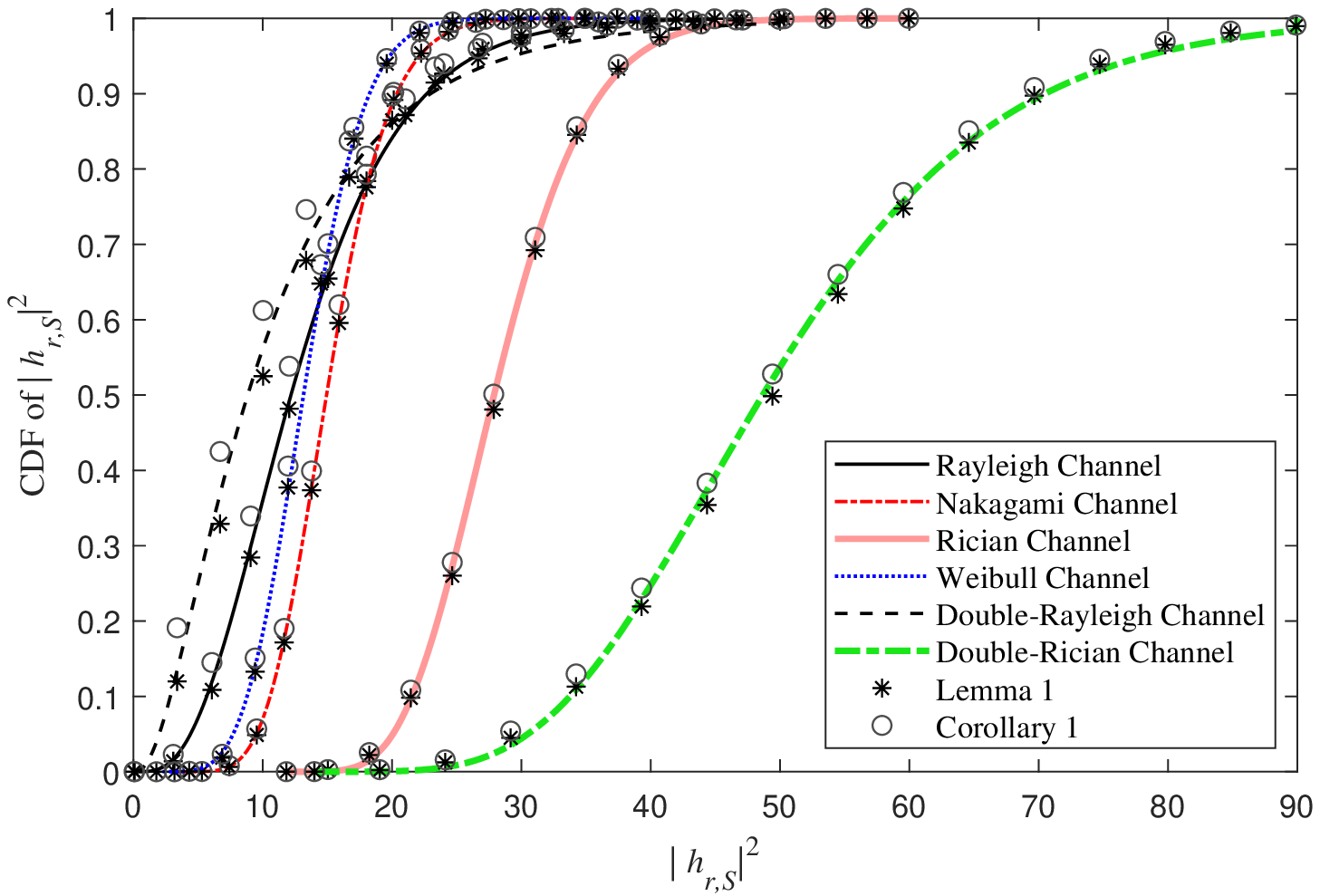}
		\label{fig_a}}
	\hfil
	\subfigure[]{\includegraphics[width=3.1in]{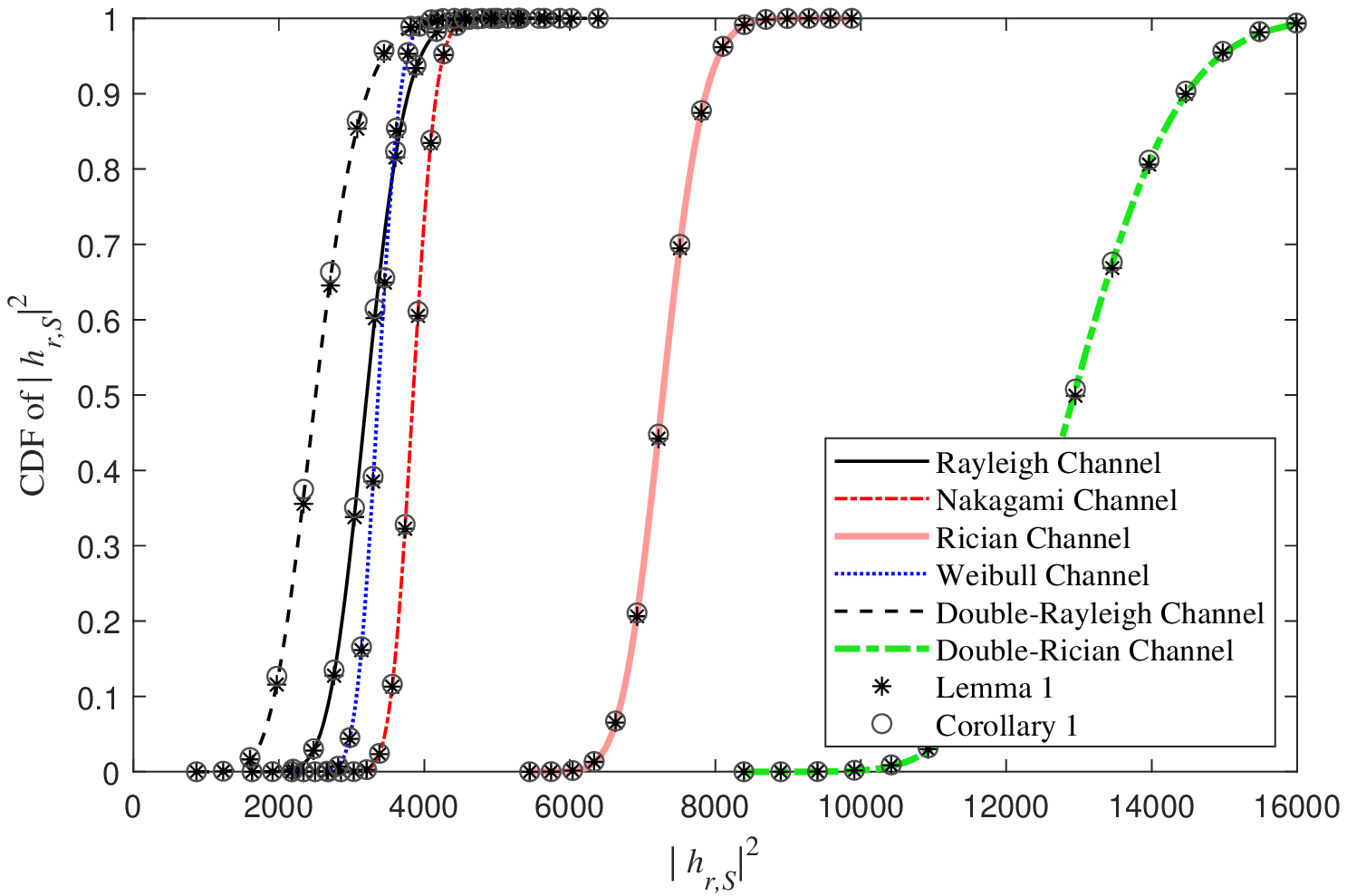}
		\label{fig_b}}
	\caption{CDF versus the value of $|h_{r,S}|^2$ with different small-scale fading models: (a) the number of STAR-RIS elements $N=4$; (b) the number of STAR-RIS elements $N=64$.}
	\label{fig: fitting}
	\vspace{-0.1 cm}
\end{figure*}
In this subsection, we pay attention to some typical small-scale fading models. Since the approximated expression of the CPG for the interference signal is similar to that in \cite{1gammaappr}, we mainly focus on the desired signal in the rest of this section. When the parameters of the particular distribution are predefined, we can easily derive the fitted Gamma distribution. For the cases $N$ is large, simple asymptotic expressions are obtained based on \textbf{Corollary \ref{corollary: gamma fitting large N}}. The results are concluded in table \ref{table: channel models}. Detailed discussions between channel parameters and analytical fitting results are as follows.
\subsubsection{Rayleigh Channel}
The channel gain of the Rayleigh channel obeys the Rayleigh distribution with the scale parameter $\delta_1>0$. We can find that the shape parameter $k_r$ of the fitted Gamma distribution is unrelated to the Rayleigh parameter $\delta_1$.

\subsubsection{Nakagami-$m$ Channel}
In this case, the channel gain obeys the Nakagami distribution with the shape parameter $m_2 \ge \frac{1}{2}$ and the scale parameter $\Omega_2 > 0$. The result shows that $k_r$ is related to $m_2$ but not related to $\Omega_2$.

\subsubsection{Rician Channel}
The deployment of RISs is expected to provide LoS transmission links, whose small-scale fading can be characterized by the Rician fading model. According to \cite{0STARRISMu}, the channel gain of the Rician channel shown as follows consists of two portions
\begin{align}
	h_{r,n} =  \sqrt{\frac{K_3}{K_3+1}} h_{r,n}^{\rm LoS} + \sqrt{\frac{1}{K_3+1}}h_{r,n}^{\rm NLoS},
\end{align}
where $K_3$ is the Rician fading factor. $h_{r,n}^{\rm LoS} = c_3$ is the deterministic LoS component. $h_{r,n}^{\rm NLoS}$ is the random NLoS component modeled as Rayleigh fading with the scale parameter $\delta_3>0$. Therefore, the Rician channel considers impacts of both LoS transmissions and NLoS transmissions.

\subsubsection{Weibull Channel}
For the Weibull channel, the channel gain obeys the Weibull distribution with the shape parameter $k_4>0$ and the scale parameter $\lambda_4 >0$. It can be found that $k_r$ is only related to the shape parameter $k_4$.

\subsubsection{Double-Rayleigh Channel}
In RIS-enabled communications, the cascaded channel is introduced. The small-scale fading correlates to channel conditions of two parts of the cascaded RIS aided link. 
The double-Rayleigh channel is the product of two independent Rayleigh channels. Similar to the result of Rayleigh channel, $k_r$ is unrelated to any Rayleigh parameter $\delta_5$ or $\delta_6$.

\subsubsection{Double-Rician Channel}
Similarly, the double-Rician channel is the product of two independent Rician channels, the channel gain of which is
\begin{align}
	h_{r,n} =  \left(\sqrt{\frac{K_7}{\hat{K}_7}} c_7 + \sqrt{\frac{1}{\hat{K}_7}}h_{7} \right)\left(\sqrt{\frac{K_8}{\hat{K}_8}} c_8 + \sqrt{\frac{1}{\hat{K}_8}}h_{8} \right),
\end{align}
where $\hat{K}_7 = K_7+1$, $\hat{K}_8 = K_8+1$, $h_7$ and $h_8$ are Rayleigh random variables with parameters $\delta_7$ and $\delta_8$, respectively.

\subsection{Fitting Accuracy}
To validate the accuracy of the proposed fitting method, we plot the cumulative distribution function (CDF) of $|h_{r,S}|^2$ in Fig. \ref{fig: fitting}. Six different small-scale fading models as we discussed in section \ref{section: case study} are considered. In Fig. \ref{fig: fitting}, we use lines to present the Monte Carlo simulation results and marks to depict the analytical fitting results. The simulation parameters are set as: $\{\delta_1,\delta_3,\delta_5,\delta_6,\delta_7, \delta_8\} = \sqrt{\frac{1}{2}}$, $\{m_2, k_4\} = 4$, $\{\Omega_2, K_3,K_7,K_8 , c_3,c_7, c_8, \lambda_4\} = 1$  \cite{0STARRISMu,1CLT2,1exact}.

We can observe that the fitted Gamma distributions shown in \textbf{Lemma \ref{lemma: Gamma fitting}} and \textbf{Corollary \ref{corollary: gamma fitting large N}} become more accurate with a larger number of STAR-RIS elements. This can be attributable to the application of CLT. For both cases $N=4$ and $N=64$, the analytical results provided in \textbf{Lemma \ref{lemma: Gamma fitting}} fits the numerical results well, while the results in \textbf{Corollary \ref{corollary: gamma fitting large N}} should be applied in large $N$ cases.

\section{Coverage Probability}
In this section, we derive the general expressions of the coverage probability for the typical UE and the connected UE. 
For comparison, the results in conventional RIS aided networks are also obtained.    

Based on the previous analysis in Section \ref{Sec: Channel}, the composite small-scale fading power for the desired signal can be fitted by Gamma distribution $\Gamma(k_r,\theta _r)$ while for the interference the Gamma distribution is $\Gamma(1,\theta _I)$. Different from the power distribution of the Nakagami-$m$ fading channel in conventional networks,  $\Gamma(k_r,\theta _r)$ is a general Gamma distribution and the value of $k_r$ might be very large. Therefore, Alzer's inequality \cite{2Alzer} employed in most existing works is no longer efficient in our scenarios. Although the Gil-Pelaez theorem is available for arbitrary distributions \cite{2Pelaez}, an extra fold of integral is introduced, which brings challenges to obtaining insights from the complex analytical expressions. Sparked by the above reasons, we provide a novel analytical method in this work.

Before deriving the performance expressions, let us introduce some preliminary definitions.
\begin{definition}
For a non-negative integer $m$, we use $\xi_m (a,b,c;x)$ to describe the following expression related to the Gauss hypergeometric function
\begin{align}
	&\xi_m(a,b,c;x) \nonumber \\ &\triangleq  \frac{(a)_{m}(-2/b)_m(-c)^m }{(1-2/b)_{m}} {}_2F_1 \left(a,-\frac{2}{b};1-\frac{2}{b};-cx \right) ,
\end{align}
where $(x)_m$ is the Pochhammer's symbol. If $m=0$, $(x)_0 = 1$; otherwise, $(x)_m = x(x+1)\cdots(x+m-1)$.
\end{definition}

\begin{definition}\label{definition: 2}
We define ${\bar k}_s$ as the nearest positive integer of $k_r$, which satisfies
\begin{align}
	{\bar k}_s = \mathop{\arg\min}_{x} |x-k_r|,x \in \mathbb{N}^+.
\end{align}

\end{definition}

\subsection{Laplace Transform of Interference}
Since the Laplace transform of the interference is the essential part of the coverage probability, we derive these expressions first.

The typical UE suffers the interference $I_t$ that consists of two portions: 1) transmissive interference from the BSs located on the back of the serving STAR-RIS; and 2) reflective interference from the BSs which are in front of the serving STAR-RIS. This Laplace transform is presented in the following lemma.
\begin{lemma}\label{lemma: Lap I_r}
In STAR-RIS aided networks, the Laplace transform of the interference for the typical UE can be derived as
\begin{align}\label{eq: Lap I_r}
	{\cal L}_{I_ {t}}(s) &=  \exp \left( - \frac{1}{2} \pi \lambda_B {r_t}^2  \left( \xi_0 \left(1,\alpha_r, \beta _{\rm T}\eta_t;s \right) -1  \right)\right) \nonumber \\
	& \times \exp \left( - \frac{1}{2} \pi \lambda_B {r_t}^2  \left( \xi_0 \left( 1,\alpha_r, \beta _{\rm R}\eta_t;s \right) -1  \right)\right),
\end{align}
where  $\eta _t = \theta_I P_B L_{t,i}^{(k)}$.
\end{lemma}
\begin{IEEEproof}
See Appendix A.
\end{IEEEproof}

For the connected UE, the interference $I_c$ is also from the STAR-RIS aided link. In the following proposition, we provide the Laplace transform of $I_c$.
\begin{proposition}
The Laplace transform of the interference for the connected UE can be given by
\begin{align}\label{eq: Lap I_d}
	{\cal L}_{I_ {c}}(s) &=  \exp \left( - \frac{1}{2} \pi \lambda_B {r_t}^2  \left( \xi_0 \left( 1,\alpha_r, \beta _{\rm T}\eta_c;s \right) -1  \right)\right) \nonumber \\
			& \times \exp \left( - \frac{1}{2} \pi \lambda_B {r_t}^2  \left( \xi_0 \left( 1,\alpha_r, \beta _{\rm R}\eta_c;s \right) -1  \right)\right),
\end{align}
where $\eta_c = \theta_{I} P_B L_{c,i}^{(k)}$.
\end{proposition}

For a fair comparison, we employ one reflecting-only RIS and one transmitting-only RIS at the same location as the assisted STAR-RIS. Both these two conventional RISs have $N/2$ elements. We denote the small-scale fading CPG for the signal and the interference of the conventional RIS aided link as $\Gamma \left( k_{r,con}, \theta_{r,con}\right)$ and $\Gamma \left( 1, \theta_{I,con}\right)$, respectively. 
\begin{lemma}\label{lemma: Lap I_r R}
In conventional RIS aided networks, the Laplace transform of the interference for the UE $u_\epsilon$ $(\epsilon \in \{t,c\})$  can be derived as
\begin{align}\label{eq: Lap I_r R}
	{\cal L}_{I_ {con}}(s) &=  \exp \left( -  \pi \lambda_B {r_t}^2  \left( \xi_0 \left( 1,\alpha_r, \eta_{\epsilon,con};s \right) -1  \right)\right),
\end{align} 
where $\eta _{t,con} = \theta_{I,con} P_B L_{\epsilon,i}^{(k)}$.
\end{lemma}
\begin{IEEEproof}
The proof is similar to \textbf{Lemma \ref{lemma: Lap I_r}} and hence we skip it here.
\end{IEEEproof}

\subsection{Coverage Performance for the Typical UE}
In this work, the coverage probability for the typical UE is defined as the probability that the typical UE can successfully transmit signals with a targeted SINR $\tau_t$. The typical UE only decodes its message after a successful SIC process. The coverage probability is expressed as
\begin{align}\label{eq: cp typical def}
	P_{t} &=  \mathbb{P} (\gamma_{t \to c} > \tau_c, \gamma_{t} > \tau_t),
\end{align}
where $\tau_c$ is the target SINR for the connected UE.

Considering the signal transmission mode at the serving STAR-RIS for the typical UE is $\chi \in \{{\rm T},{\rm R}\}$, the conditional coverage probability of the typical UE can be rewritten as
\begin{align}
	P_{t,STAR,\chi}^{}|_{d_{t,i}^{(k)}} &=  \mathbb{P} \left( |h_r|^2 > \tau_t^* \frac{I_t + {n_0}^2}{\beta_\chi P_BL_{t,i}^{(k)}}   \right),
\end{align}
where $\tau_t^* = \max \left( \frac{\tau_c}{a_c-\tau_ca_t},\frac{\tau_t}{a_t}\right)$.

Utilizing the scaling feature of the Gamma distribution, we have $\frac{1}{\theta_r} |h_r|^2 \sim \Gamma(k_r,1)$. However, $k_r$ is not an integer in most cases. For tractability, we introduce ${\bar k}_r$ defined in \textbf{Definition \ref{definition: 2}} to deduce the analytical coverage expressions.

\begin{theorem}\label{theorem: cp typical}
In STAR-RIS aided networks, the coverage probability for the typical UE is derived as
\begin{align}
	P_{t,STAR} &= \int_{0}^\infty  \int_{0}^\infty  \pi \lambda_B r_1 f_{RU}(r_2) \sum_{m=0}^{{\bar k}_r-1}\frac{(-1)^m e^ {V_{\rm T}^{(0)}(1,1)}}{m!} \nonumber \\&  \times {\cal B}_m \left( V_{\rm T}^{(1)}(1,1), ..., V_{\rm T}^{(m)}(1,1) \right) dr_1 dr_2 \nonumber\\
	& + \int_{0}^\infty  \int_{0}^\infty \pi \lambda_B r_1 f_{RU}(r_2) \sum_{m=0}^{{\bar k}_r-1}\frac{(-1)^m  e ^ {V_{\rm R}^{(0)}(1,1)}}{m!} \nonumber \\&  \times {\cal B}_m \left( V_{\rm R}^{(1)}(1,1), ..., V_{\rm R}^{(m)}(1,1) \right) dr_1 dr_2,
\end{align}
with
\begin{align}
	V_{\chi}^{(m)}(z,x) =& -\Delta_{t,\chi}^{(m)} - \frac{1}{2}\pi {r_1}^2\lambda_B
	\xi_m \left( 1,\alpha_r, zD_{{\rm T}\chi}; x \right) \nonumber \\&  - \frac{1}{2}\pi {r_1}^2\lambda_B\xi_m \left( 1, \alpha_r, zD_{{\rm R}\chi}; x \right),
\end{align}
where ${\cal B}_m \left(x_1, ..., x_m \right)$ is the $m$th complete Bell polynomial. $\Delta_{t,\chi}^{(0)} = s_{t,\chi }{n_0}^2x$, $\Delta_{t,\chi}^{(1)} = s_{t,\chi }{n_0}^2$, and $\Delta_{t,\chi}^{(m)} = 0$ when $m \ge 2$. $D_{{\rm T}\chi} = \frac{\theta_I\tau_t^*\beta _{\rm T}}{\theta_r\beta _\chi}$, $D_{{\rm R}\chi} = \frac{\theta_I\tau_t^*\beta _{\rm R}}{\theta_r\beta _\chi}$, and $s_{t,\chi } = \frac{\tau_t^* (r_1r_2)^{\alpha_r}}{\theta_r \beta_\chi P_B C_r } $.
\end{theorem}
\begin{IEEEproof}
	See Appendix B.
\end{IEEEproof}

In STAR-RIS aided communications, the operating parameters make difference to the system performance. Thus, the following corollary provides the optimal energy splitting coefficient in terms of coverage performance.
\begin{corollary}\label{corollary: opt beta}
When $\beta_{\rm T} = \beta_{\rm R} = \frac{1}{2}$, the maximum coverage probability of the typical UE is
\begin{align}
	P_{t,STAR}^{\rm max} &= \int_{0}^\infty \int_{0}^\infty 2\pi \lambda_B r_1 f_{RU}(r_2) \sum_{m=0}^{{\bar k}_r-1}\frac{(-1)^m e^{{\tilde V}^{(0)}(1,1)}}{m!} \nonumber \\&  \times  {\cal B}_m \left( {\tilde V}^{(1)}(1,1), ..., {\tilde V}^{(m)}(1,1) \right) dr_1 dr_2,
\end{align}
where ${\tilde V}^{(m)}(z,x) = -\Delta_{t,max}^{(m)}  - \pi {r_1}^2\lambda_B \xi_m \left( 1,\alpha_r, \frac{z\theta_I\tau_t^*}{\theta_r}; x \right)$, $s_{t,max } = \frac{2\tau_t^* (r_1r_2)^{\alpha_r}}{\theta_r  P_B C_r } $, $\Delta_{t,max}^{(0)} = s_{t,max }{n_0}^2x$, $\Delta_{t,max}^{(1)} = s_{t,max }{n_0}^2$, and $\Delta_{t,max}^{(m)} = 0$ when $m \ge 2$.
\end{corollary}

\begin{IEEEproof}
	See Appendix C.
\end{IEEEproof}

\begin{remark}
The results obtained in \emph{\textbf{Corollary \ref{corollary: opt beta}}} can be explained that $\beta_{\rm T} = \beta_{\rm R} = \frac{1}{2}$ guarantees the typical UE to receive the strongest signal from its serving BS instead of other BSs under random scenarios. 
When the assisted mode for the typical UE is predefined, STAR-RISs have the capability of controlling the received signal power by adjusting the energy splitting coefficient and hence are able to meet various QoS requirements of UEs at different sides. 
\end{remark}

\begin{proposition}\label{theorem: cp typical R}
In conventional RIS aided networks, the coverage probability for the typical UE is derived as
\begin{align}
	P_{t,con} &= \int_{0}^\infty \!\int_{0}^\infty 2\pi \lambda_B r_1 f_{RU}(r_2) \sum_{m=0}^{{\bar k}_{r,con}-1}\frac{(-1)^m e^{V_{con}^{(0)}(1,1)}}{m!} \nonumber \\&  \times {\cal B}_m \left( V_{con}^{(1)}(1,1), ..., V_{con}^{(m)}(1,1) \right) dr_1 dr_2,
\end{align}
with
\begin{align}
	V_{con}^{(m)}(x) =& -\Delta _{t,con} ^{(m)}  - \pi {r_1}^2\lambda_B
	\xi_m \left( 1,\alpha_r, \frac{\theta_I\tau_t^*}{\theta_r}; x \right) ,
\end{align}
where $s_{t,con} = \frac{\tau_t^* (r_1r_2)^{\alpha_r}}{\theta_{r,con}  P_B C_r } $, $\Delta_{con}^{(0)} = s_{t,con}{n_0}^2x$, $\Delta_{t,con}^{(1)} = s_{t,con}{n_0}^2$, and $\Delta_{t,con}^{(m)} = 0$ when $m \ge 2$.
\end{proposition}
\begin{IEEEproof}
Since it is of the same probability for the typical UE to associate with a reflecting-only RIS or a transmitting-only RIS, we only need to derive the coverage probability when the typical UE associates with the reflecting-only RIS and then double the result. In this case, the PDF of the serving distance between the BS and the assistant RIS is
\begin{align}
	f_{BR}^{ref}(x) =& \pi \lambda_b x \exp \left( -\frac{1}{2}\lambda_b x^2 \right).
\end{align}
Then, using the similar proof in \textbf{\textbf{Theorem \ref{theorem: cp typical}}} and \eqref{eq: Lap I_r R}, this theorem can be proved.	
\end{IEEEproof}

Now let us consider the interference-limited case as a special case. In this case, the noise is negligible compared to the interference, i.e., $I_r \gg {n_0}^2$, so we focus on the SIR coverage. We can obtain closed-form expressions for the coverage probability of the typical UE shown as the following corollaries.
\begin{corollary}\label{corollary: cp typical closed}
When $I_r \gg {n_0}^2$, the coverage probability of the typical UE in STAR-RIS aided networks can be expressed in a closed form as follows
	\begin{align}\label{eq: cp closed}
		P_{t,STAR} &= \sum_{m=0}^{{\bar k}_r-1} \frac{(-1)^m}{m!} \sum_{l=1}^{m} \frac{(-1)^{l}l!}{\left( V_{SIR,\rm T}^{(0)}(1,1) \right)^{l+1}} \nonumber \\&  \times   {\cal B}_{m,l} \left( V_{SIR,\rm T}^{(1)}(1,1), ..., V_{SIR,\rm T}^{(m-l+1)}(1,1) \right) \nonumber \\
		&+ \sum_{m=0}^{{\bar k}_r-1} \frac{(-1)^m}{m!} \sum_{l=1}^{m} \frac{(-1)^{l}l!}{\left( V_{SIR,\rm R}^{(0)}(1,1) \right)^{l+1}}  \nonumber \\&  \times {\cal B}_{m,l} \left( V_{SIR,\rm R}^{(1)}(1,1), ..., V_{SIR,\rm R}^{(m-l+1)}(1,1) \right),
	\end{align}
	where $V_{SIR,\chi}^{(m)}(z,x) = 
	\xi_m \left( 1,\alpha_r, zD_{{\rm T}\chi}; x \right) + \xi_m \left( 1,\alpha_r, zD_{{\rm R}\chi}; x \right)$.
\end{corollary}
\begin{IEEEproof}
Since the operators of integral and differentiation are interchangeable, we can calculate the second-order derivative of the conditional coverage probability for $\chi \in \{{\rm T}, {\rm R}\}$
\begin{align}
	&P_{t,STAR,\chi} = \sum_{m=0}^{{\bar k}_r-1}\frac{(-1)^m}{m!} \nonumber \\&  \times \left[ \frac{\partial ^m}{ \partial x^m} \frac{2}{\xi_0 \left( 1,\alpha_r, D_{{\rm T}\chi}; x \right) + \xi_0 \left( 1,\alpha_r, D_{{\rm R}\chi}; x \right)} \right]_{x=1}.
\end{align}

Then we recall Fa\`a di Bruno's formula as we have stated in the proof of \textbf{Theorem \ref{theorem: cp typical}}, this corollary is proved.
\end{IEEEproof}

\begin{corollary}\label{corollary: cp typical closed R}
When $I_r \gg {n_0}^2$, the coverage probability of the typical UE in conventional RIS aided networks can be expressed in a closed form as follows
\begin{align}\label{eq: cp closed R}
	P_{t,con} &= \sum_{m=0}^{{\bar k}_{r,con}-1} \frac{(-1)^m}{m!} \sum_{l=1}^{m} \frac{(-1)^{l}l!}{\left( V_{SIR,con}^{(0)}(1,1) \right)^{l+1}}   \nonumber \\&  \times {\cal B}_{m,l} \left( V_{SIR,con}^{(1)}(1,1), ..., V_{SIR,con}^{(m-l+1)}(1,1) \right) ,
\end{align}
where $V_{SIR,con}^{(m)}(z,x) = \xi_m \left( 1,\alpha_r, \frac{z\theta_I\tau_t^*}{\theta_r}; x \right)$.
\end{corollary}
\begin{IEEEproof}
The poof is similar to \textbf{Corollary \ref{corollary: cp typical closed}}.
\end{IEEEproof}

%\begin{remark}
%By observing \eqref{eq: cp closed} and \eqref{eq: cp closed R} we can find that when $\beta^{\rm T} = \beta^{\rm R} = \frac{1}{2}$, the value of coverage probability in two kinds of RIS aided networks are equal. Therefore, \eqref{eq: cp closed R} is the tight upper bound for \eqref{eq: cp closed} in interference-limited scenarios.
%\end{remark}

\subsection{Coverage Performance for the Connected UE}
The connected UE decodes its own message by treating the typical UE as noise, so the coverage probability is
\begin{align}\label{eq: cp connected def}
	P_{c} &=  \mathbb{P} (\gamma_{c} > \tau_c).
\end{align}

Based on \eqref{eq: SINR connected}, the coverage probability of the connected UE can be rewritten as
\begin{align}
	P_{c}|_{d_{c,i}^{(k)} } &=  \mathbb{P} \left( |h_c|^2 > \tau_c^* \frac{I_c + {n_0}^2}{P_BL_{c,i}^{(k)}}   \right),
\end{align}
where $\tau_c^* = \frac{\tau_c}{a_c - a_t\tau_c}$.

%\vspace{-0.8 cm}
Similar to the typical UE, the connected UE associates to the BS assisted by the STAR-RIS. We can easily obtain the exact analytical expression shown in the following theorem.

\begin{theorem}
In STAR-RIS aided networks, the coverage probability for the connected UE can be given by
\begin{align}
	P_{c,STAR} =& \int_{0}^\infty  \pi \lambda_B r_1 \sum_{m=0}^{{\bar k}_{r}-1}\frac{(-1)^m e^ {\Lambda_{\rm T}^{(0)}(1,1)}}{m!} \nonumber \\&  \times {\cal B}_m \left( \Lambda_{\rm T}^{(1)}(1,1), ..., \Lambda_{\rm T}^{(m)}(1,1) \right) dr_1  \nonumber\\
	& + \int_{0}^\infty \pi \lambda_B r_1 \sum_{m=0}^{{\bar k}_r-1}\frac{(-1)^m  e ^ {\Lambda_{\rm R}^{(0)}(1,1)}}{m!} \nonumber \\&  \times {\cal B}_m \left( \Lambda_{\rm R}^{(1)}(1,1), ..., \Lambda_{\rm R}^{(m)}(1,1) \right) dr_1,
\end{align}
with
\begin{align}
	\Lambda_{\chi}^{(m)}(z,x) =& -\Delta_{c,\chi}^{(m)}  - \frac{1}{2}\pi {r_1}^2\lambda_B
	\xi_m \left( 1,\alpha_r, zD_{{\rm T}\chi}; x \right)  \nonumber \\&  - \frac{1}{2}\pi {r_1}^2\lambda_B\xi_m \left( 1,\alpha_r, zD_{{\rm R}\chi}; x \right) ,
\end{align}
where $s_{c,\chi } = \frac{\tau_c^* (r_1d_c)^{\alpha_r}}{\theta_r \beta_\chi P_B C_r } $. $\Delta_{c,\chi}^{(0)} = s_{c,\chi}{n_0}^2x$, $\Delta_{c,\chi}^{(1)} = s_{c,\chi}{n_0}^2$ and $\Delta_{c,\chi}^{(m)} = 0$ when $m \ge 2$.
\end{theorem}
\begin{IEEEproof}
Utilizing the fact that $d_{c}^{(k)}=d_c$ is a constant and the proof in \textbf{Theorem \ref{theorem: cp typical}}, this theorem can be proved.
\end{IEEEproof}
\begin{corollary}\label{corollary: opt beta c}
When $\beta_{\rm T} = \beta_{\rm R} = \frac{1}{2}$, the maximum coverage probability of the connected UE is
\begin{align}
	P_{c,STAR}^{\rm max} &= \int_{0}^\infty  2\pi \lambda_B r_1  \sum_{m=0}^{{\bar k}_r-1}\frac{(-1)^m e^{{\tilde \Lambda}^{(0)}(1,1)}}{m!} \nonumber \\&  \times {\cal B}_m \left( {\tilde \Lambda}^{(1)}(1,1), ..., {\tilde \Lambda}^{(m)}(1,1) \right) dr_1,
\end{align}
where ${\tilde \Lambda}^{(m)}(z,x) = -\Delta_{c,max}^{(m)}  - \pi {r_1}^2\lambda_B \xi_m \left( 1,\alpha_r, \frac{z\theta_I\tau_c^*}{\theta_r}; x \right)$, $s_{c,max } = \frac{2\tau_c^* (r_1d_c)^{\alpha_r}}{\theta_r  P_B C_r } $, $\Delta_{c,max}^{(0)} = s_{c,max }{n_0}^2x$, $\Delta_{c,max}^{(1)} = s_{c,max }{n_0}^2$, and $\Delta_{c,max}^{(m)} = 0$ when $m \ge 2$.
\end{corollary}
\begin{IEEEproof}
The proof is as same as \textbf{Corollary \ref{corollary: opt beta}}.
\end{IEEEproof}

{\begin{remark}
When $\beta_{\rm T} = \beta_{\rm R} = \frac{1}{2}$ both the typical UE and the connected UE achieve the maximum coverage probability.
This illustrates that considering randomly deployed networks, the average system coverage can be optimized by adjusting the energy splitting coefficient of STAR-RISs.
\end{remark}

We also obtain the coverage probability in conventional RIS aided networks showing as follows.

\begin{proposition}\label{prop: cp connect R}
	In conventional RIS aided networks, the coverage probability for the connected UE is derived as
	\begin{align}
		P_{c,con} &= \int_{0}^\infty 2\pi \lambda_B r_1  \sum_{m=0}^{{\bar k}_{r,con}-1}\frac{(-1)^m e ^{\Lambda_{con}^{(0)}(1,1)}}{m!} \nonumber \\& \times {\cal B}_m \left( \Lambda_{con}^{(1)}(1,1), ..., \Lambda_{con}^{(m)}(1,1) \right) dr_1,
	\end{align}
	with
	\begin{align}
		\Lambda_{con}^{(m)}(z,x) =& -\Delta _{c,con} ^{(m)}  - \pi {r_1}^2\lambda_B
		\xi_m \left( 1,\alpha_r, \frac{z\theta_I\tau_c^*}{\theta_r}; x \right) ,
	\end{align}
	where $s_{c,con} = \frac{\tau_c^* (r_1d_c)^{\alpha_r}}{\theta_{r,con}  P_B C_r } $, $\Delta_{c,con}^{(0)} = s_{c,con}{n_0}^2x$, $\Delta_{c,con}^{(1)} = s_{c,con}{n_0}^2$, and $\Delta_{c,con}^{(m)} = 0$ when $m \ge 2$.
\end{proposition}

%When considering the interfernce limited scenario, a closed-form expression of the coverage probability can be obtained.

\section{Ergodic Rate}
Rather than calculating the coverage probability with a predefined threshold, the ergodic rate of the STAR-RIS aided NOMA networks is determined by random channel conditions of UEs. Hence the ergodic rate can be an important metric to characterize the system performance. In this section, we evaluate the ergodic rates for both the typical UE and the connected UE. Besides, the results in conventional RIS aided networks are also obtained.

\subsection{Ergodic Rate for the Typical UE}
According to our assumption, the SIC procedure always occurs at the typical UE. If the typical UE fails to process the SIC, it can never decode its message, and hence its ergodic rate is zero. Therefore, the ergodic rate of the typical UE can be expressed as
\begin{align}\label{eq: rate typical def}
	R_{t} &=  \mathbb{E} \left[ \log_2 \left( 1  + \gamma_t \right), \gamma_{t \to c} > \tau_c \right].
\end{align}

Based on the expressions of the coverage probability, we can obtain the exact ergodic rate in two kinds of RIS aided networks in \textbf{Theorem \ref{theorem: rate typical}} and \textbf{Proposition \ref{theorem: rate typical R}}.
\vspace{-0.2 cm}
\begin{theorem}\label{theorem: rate typical}
In STAR-RIS aided networks, the ergodic rate for the typical UE is derived as
\begin{align}\label{eq: rate typical}
	R_{t,STAR} &=  \frac{1}{\ln 2} \int_{a_t\tau_c^*}^\infty \frac{{\bar F}_{t,STAR} (z)}{1+z}dz \nonumber \\& + \log_2(1+a_t\tau_c^*) {\bar F}_{t,STAR} (a_t\tau_c^*),
\end{align}
where ${\bar F}_{t,STAR} (z)$ is given by
\begin{align}
	&{\bar F}_{t,STAR} (z) = \int_{0}^\infty  \int_{0}^\infty  \pi \lambda_B r_1 f_{RU}(r_2) \sum_{m=0}^{{\bar k}_r-1}\frac{(-1)^m }{m!}  \nonumber \\& \times e^ {V_{\rm T}^{(0)}(z,1)} {\cal B}_m \left( V_{\rm T}^{(1)}(z,1), ..., V_{\rm T}^{(m)}(z,1) \right) dr_1 dr_2 \nonumber\\
	& + \int_{0}^\infty  \int_{0}^\infty \pi \lambda_B r_1 f_{RU}(r_2) \sum_{m=0}^{{\bar k}_r-1}\frac{(-1)^m }{m!}  \nonumber \\& \times e ^ {V_{\rm R}^{(0)}(z,1)} {\cal B}_m \left( V_{\rm R}^{(1)}(z,1), ..., V_{\rm R}^{(m)}(z,1) \right) dr_1 dr_2.
\end{align}
\end{theorem}
\begin{IEEEproof}
	See Appendix D.
\end{IEEEproof}

We also investigate the impact of the energy splitting coefficient on the ergodic rate in the following corollary.
\begin{corollary}\label{corollary: opt beta rate}
When $\beta_{\rm T} = \beta_{\rm R} = \frac{1}{2}$, the maximum ergodic rate of the typical UE is
\begin{align}
	R_{t,STAR}^{\rm max} &= \frac{1}{\ln 2} \int_{a_t\tau_c^*}^\infty \frac{{\bar F} _{t,STAR} ^{\rm max} (z)}{1+z}dz  \nonumber \\&  + \log_2(1+a_t\tau_c^*) {\bar F} _{t,STAR} ^{\rm max} (a_t\tau_c^*),
\end{align}
where ${\bar F}_{t,STAR}^{\rm max}(z)$ is expressed as
\begin{align}
	&{\bar F}_{t,STAR}^{\rm max}(z) =  \int_{0}^\infty \!\! \int_{0}^\infty 2\pi \lambda_B r_1 f_{RU}(r_2)   \sum_{m=0}^{{\bar k}_r-1}\frac{(-1)^m }{m!} \nonumber \\& \times e^{{\tilde V}^{(0)}(z,1)} {\cal B}_m \left( {\tilde V}^{(1)}(z,1), ..., {\tilde V}^{(m)}(z,1) \right) dr_1 dr_2.
\end{align}

\end{corollary}
\begin{IEEEproof}
We denote $\beta = \beta _{\rm T} = 1- \beta_{\rm R}$. Following the similar procedure of the proof in \textbf{Corollary \ref{corollary: opt beta}}, we can calculate $\frac{\partial}{\partial \beta} R_{t,STAR}^{\rm max}$ and find that $\frac{\partial} {\partial \beta} R_{t,STAR}^{\rm max} = 0$ when $\beta = \frac{1}{2}$. Then the maximum ergodic rate is obtained.
\end{IEEEproof}

\begin{proposition}\label{theorem: rate typical R}
In conventional RIS aided networks, the ergodic rate for the typical UE is
\begin{align}\label{eq: rate typical R}
	R_{t,con} &= \frac{1}{\ln 2} \int_{a_t\tau_c^*}^\infty \frac{{\bar F}_{t,con} (z)}{1+z}dz \nonumber \\& + \log_2(1+a_t\tau_c^*) {\bar F}_{t,con} (a_t\tau_c^*),
\end{align}
where ${\bar F}_{t,con} (z)$ is expressed as
\begin{align}
	&{\bar F}_{t,con} (z) = \int_{0}^\infty \int_{0}^\infty 2\pi \lambda_B r_1 f_{RU}(r_2) \sum_{m=0}^{{\bar k}_{t,con}-1}\frac{(-1)^m }{m!} \nonumber \\& \times e^{V_{con}^{(0)}(z,1)} {\cal B}_m \left( V_{con}^{(1)}(z,1), ..., V_{con}^{(m)}(z,1) \right) dr_1 dr_2.
\end{align}
\end{proposition}
%\begin{IEEEproof}
%	The proof is similar to \textbf{Theorem \ref{theorem: rate typical}}.
%\end{IEEEproof}
%\begin{remark}
%Note that ${\bar F}_t (z) = P _{cov,t} (\frac{z}{a_t})$ and ${\bar F}_t^{ref} (z) = P _{cov,t} ^{ref} (\frac{z}{a_t})$, the expressions of the ergodic rate come from the coverage probability. Therefore, we can obtain the similar conclusion as in coverage probability. Based on \eqref{eq: cp closed}, \eqref{eq: cp closed R}, and Corollary \ref{corollary: opt beta rate}, we can find that when $I_r \gg {n_0}^2$, the ergodic rate in traditional RIS aided networks is the tight upper bound of that in STAR-RIS aided scenarios.
%\end{remark}

\subsection{Ergodic Rate for the Connected UE}
For the connected UE, the ergodic rate can be expressed as
\begin{align}\label{eq: rate connected def}
	R_{c} &=  \mathbb{E} \left[ \log_2 \left( 1  + \gamma_c \right) \right].
\end{align}

We first provide the exact expression of the ergodic rate in STAR-RIS aided networks.
\begin{theorem}\label{theorem: rate connected}
In STAR-RIS aided networks, the ergodic rate for the connected UE is derived as
	\begin{align}\label{eq: rate connected}
		R_{c,STAR} &=  \frac{1}{\ln 2} \int_{0}^{\frac{a_c}{a_t}} \frac{{\bar F}_{c,STAR} (z)}{1+z}dz,
	\end{align}
	where ${\bar F}_{c,STAR} (z)$ is given by
	\begin{align}
		{\bar F}_{c,STAR} (z) &= \int_{0}^\infty  \pi \lambda_B r_1 \sum_{m=0}^{{\bar k}_{r}-1}\frac{(-1)^m e^ {\Lambda_{\rm T}^{(0)}(z,1)}}{m!}  \nonumber \\& \times {\cal B}_m \left( \Lambda_{\rm T}^{(1)}(z,1), ..., \Lambda_{\rm T}^{(m)}(z,1) \right) dr_1  \nonumber\\
		& + \int_{0}^\infty \pi \lambda_B r_1 \sum_{m=0}^{{\bar k}_r-1}\frac{(-1)^m  e ^ {\Lambda_{\rm R}^{(0)}(z,1)}}{m!}  \nonumber \\& \times {\cal B}_m \left( \Lambda_{\rm R}^{(1)}(z,1), ..., \Lambda_{\rm R}^{(m)}(z,1) \right) dr_1,
	\end{align}
\end{theorem}
\begin{IEEEproof}
The complementary cumulative distribution function (CCDF) of the decoding SINR for the connected UE is denoted as ${\bar F}_{c,STAR}(z)$, which can be expressed as
\begin{align}
	{\bar F}(z) &= \mathbb{P} \left( (a_c - a_tz) |h_c|^2 >  \frac{ \left( I_c + {n_0}^2 \right)z }{P_BC_r L_{c,i}^{(k)} }   \right) .
\end{align} 
Note that for the case $z \ge \frac{a_c}{a_t}$, ${\bar F}_c(z) = 0$ always holds. For the case $z < \frac{a_c}{a_t}$, we can obtain \eqref{eq: rate connected} by using the similar proof in \textbf{Theorem \ref{theorem: rate typical}}.
\end{IEEEproof}

Similarly, we obtain the optimal ergodic rate for the connected UE as in prior analysis.
\begin{corollary}\label{corollary: opt beta rate c}
When $\beta_{\rm T} = \beta_{\rm R} = \frac{1}{2}$, the maximum ergodic rate of the connected UE is
\begin{align}
	R_{c,STAR}^{\rm max} = \frac{1}{\ln 2} \int_{a_t\tau_c^*}^\infty \frac{{\bar F} _{c,STAR} ^{\rm max} (z)}{1+z}dz,
\end{align}
	where ${\bar F}_{c,STAR}^{\rm max}(z)$ is 
	\begin{align}
		{\bar F}_{c,STAR}^{\rm max}(z) & =  \int_{0}^\infty 2\pi \lambda_B r_1   \sum_{m=0}^{{\bar k}_r-1}\frac{(-1)^m e^{{\tilde \Lambda}^{(0)}(z,1)}}{m!}  \nonumber \\& \times {\cal B}_m \left( {\tilde \Lambda}^{(1)}(z,1), ..., {\tilde \Lambda}^{(m)}(z,1) \right) dr_1.
	\end{align}
\end{corollary}

\begin{remark}
Similar to the results in coverage probability, when $\beta^{\rm T} = \beta^{\rm R} = \frac{1}{2}$, the ergodic rate of both the typical UE and the connected UE is maximized. Thus, the appropriate energy splitting coefficient of STAR-RISs also helps to improve the achievable ergodic rate in randomly deployed networks. 
\end{remark}

\begin{proposition}
In conventional RIS aided networks, the ergodic rate for the connected UE is
	\begin{align}
		R_{c,con} &=  \frac{1}{\ln 2} \int_0^{\frac{a_c}{a_t}} \frac{{\bar F}_{c,con} (z)}{1+z}dz,
	\end{align}
	with
	\begin{align}
		{\bar F}_{c,con} (z) &= \int_{0}^\infty 2\pi \lambda_B r_1 \sum_{m=0}^{{\bar k}_{t,con}-1}\frac{(-1)^m e^{\Lambda_{con}^{(0)}(z,1)}}{m!} \nonumber \\& \times {\cal B}_m \left( \Lambda_{con}^{(1)}(z,1), ..., \Lambda_{con}^{(m)}(z,1) \right) dr_1.
	\end{align}
\end{proposition}

\section{Numerical Results}
\begin{figure*}[t!] 
	\centering
	\subfigure[]{\includegraphics[width=3.1 in]{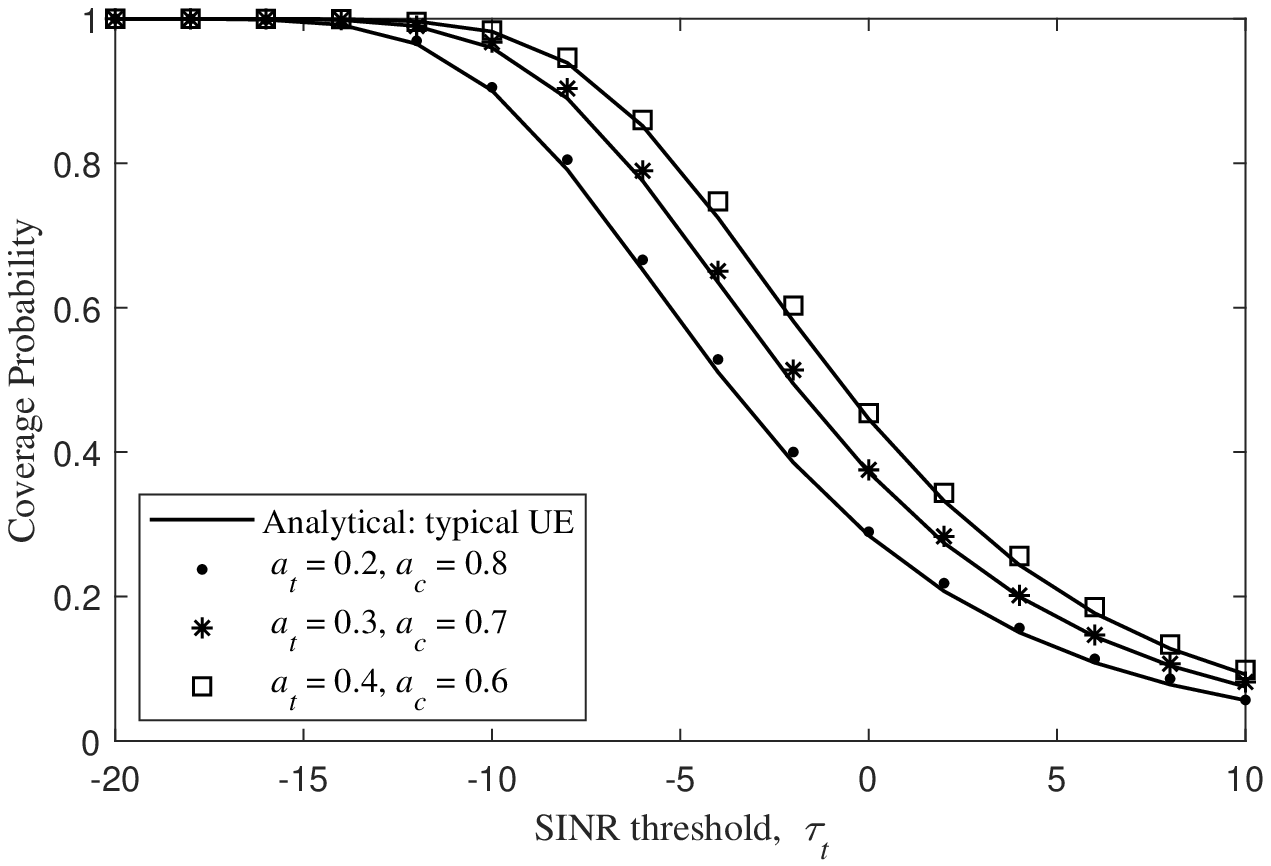}
		\label{v: fig_a}}
	\hfil
	\subfigure[]{\includegraphics[width=3.1 in]{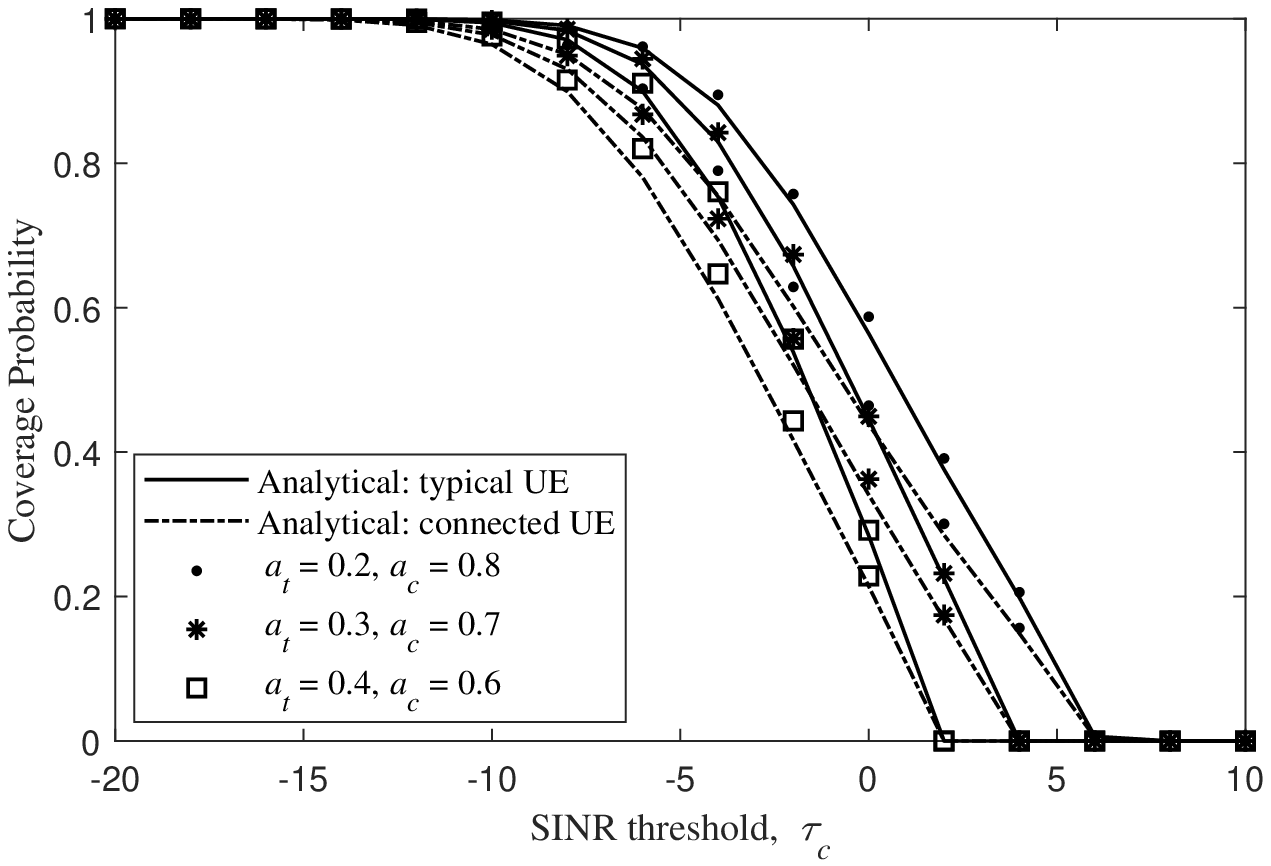}
		\label{v: fig_b}}
	\caption{Coverage probability versus SINR threshold with $a_t \in \{0.2,0.3,0.4\}$, $N=4$, $\lambda_B = 4\lambda_b$, and $\lambda_R = 20\lambda_r$: (a) the target SINR for the typical UE $\tau_t$; (b) the target SINR for the connected UE $\tau_c$.}
	\label{fig: verification}
\end{figure*}
\begin{figure*}[t!] 
	\centering
	\subfigure[]{\includegraphics[width=3.1in]{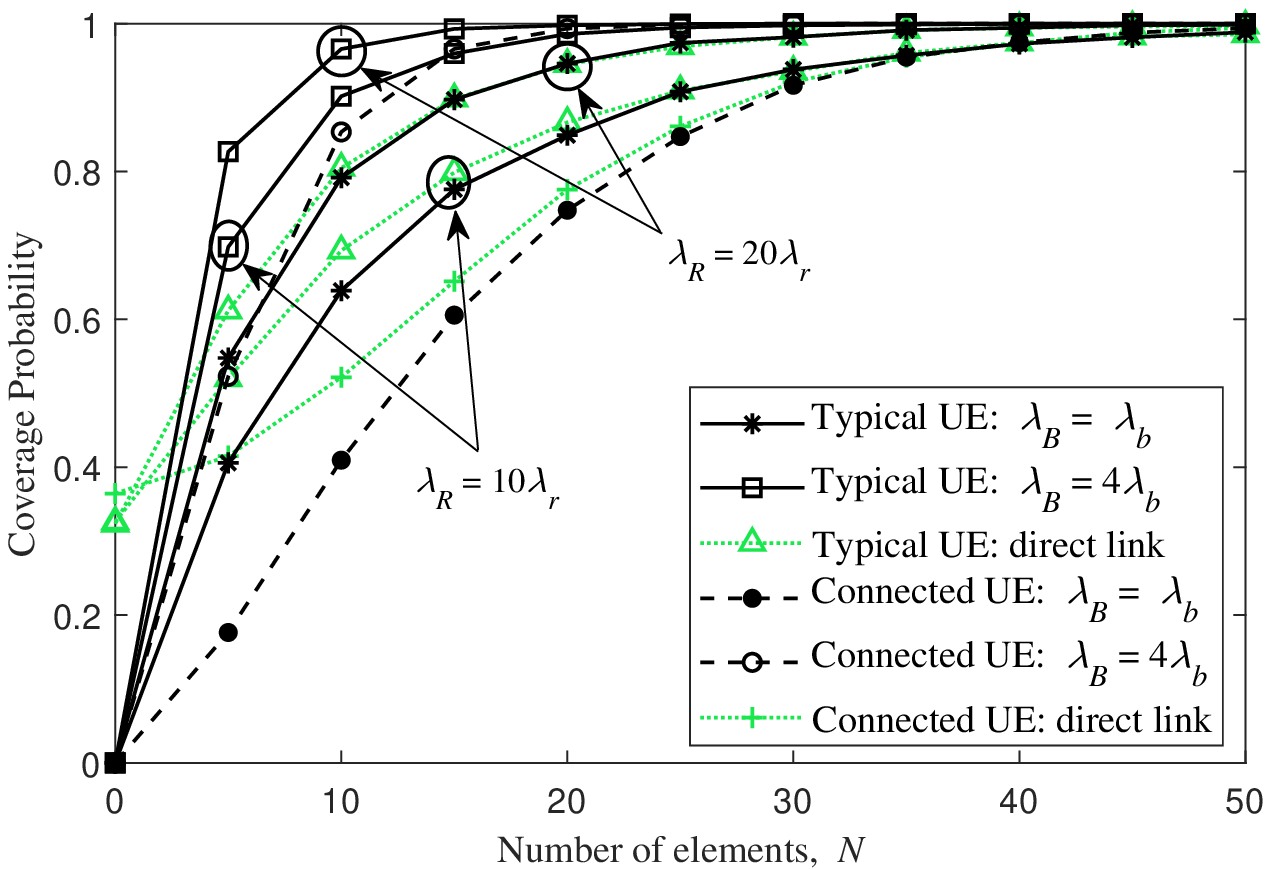}
		\label{n: fig_a}}
	\hfil
	\subfigure[]{\includegraphics[width=3.1in]{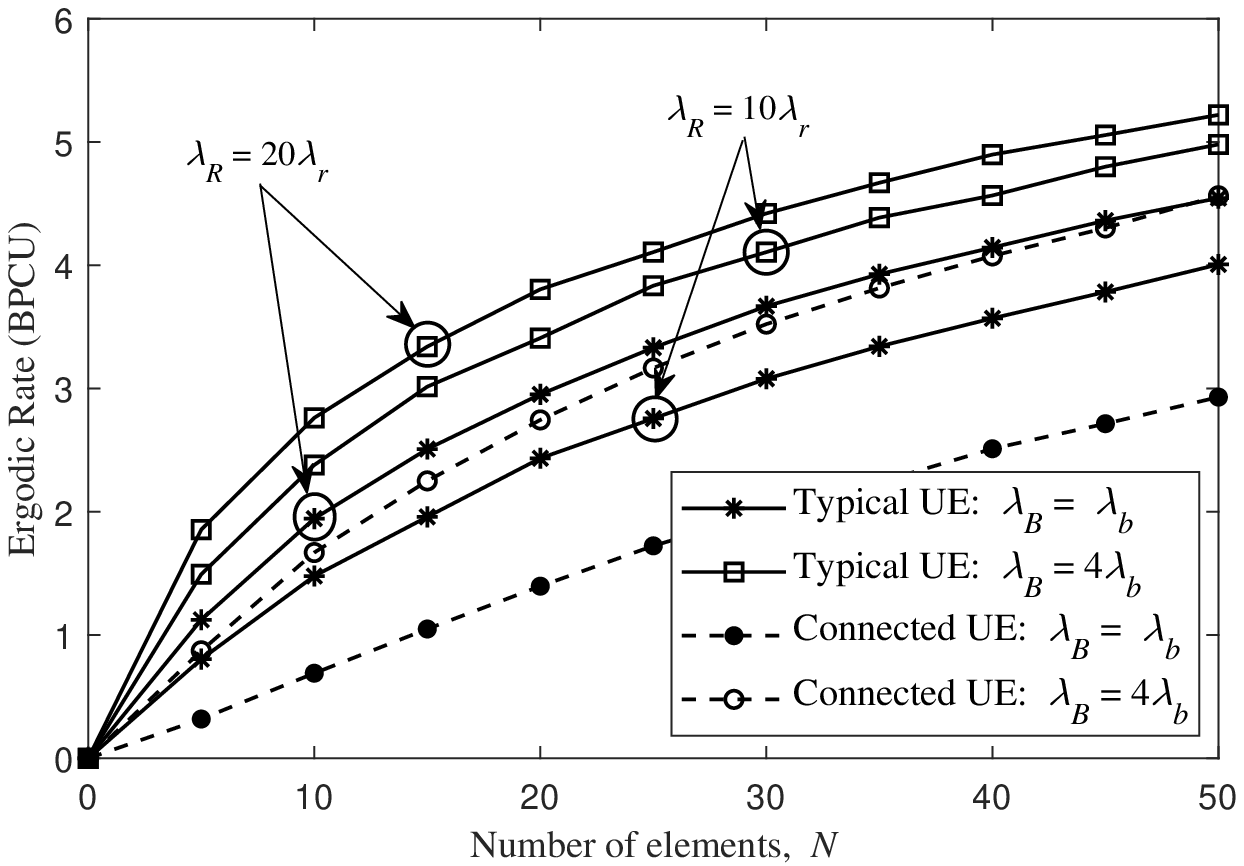}
		\label{n: fig_b}}
	\caption{System performance versus the number of STAR-RIS elements: (a) coverage probability; (b) ergodic rate.
	}
	\label{fig: element}
\end{figure*}
\begin{figure} [t!]
	\centering
	\includegraphics[width = 3.1 in] {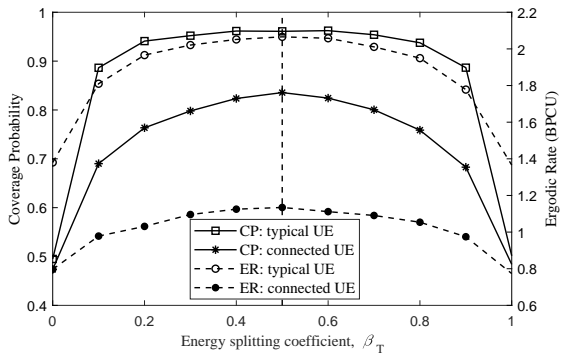}
	\caption{System performance versus energy splitting coefficient $\beta_{\rm T}$ with $N=16$, $\lambda_B=\lambda_b$, and $\lambda_R= 10\lambda_r$. "CP" represents coverage probability and "ER" represents ergodic rate.}
	\label{fig: energy splitting}
\end{figure}
In this section, we first present the numerical results to verify our analytical expressions derived in previous sections and then provide some interesting insights. For the small-scale fading model, we mainly focus on the double-Rician fading channel as we discussed in Section III. The noise power is ${n_0}^2 = -174 + 10\log_{10}W$, where $W$ is the bandwidth. We denote $\lambda_b = 2$ km$^{-2}$ and $\lambda_r = 10$ km$^{-2}$ as the BS and RIS reference densities, respectively. Without otherwise stated, the simulation parameters are defined as follows. The transmit power $P_B$ is 30 dBm. The path loss exponent is $\alpha_r = 2.8$. The intercept is $C_r = -30$ dB. The energy splitting coefficients are $\beta_{\rm T} = \beta_{\rm R} = 0.5$. The target rate is set to be equal as $\rho_t = \rho_c = 0.1$ bit per channel use (BPCU) for the typical UE and the connected UE. Thus, the target SINR are $\tau _t = 2^{\rho_t} -1$ and $\tau _c = 2^{\rho_c} -1$. The power allocation coefficients are $a_t = 0.4$ and $a_c = 0.6$. The bandwidth is $W=100$ MHz. The distance between the connected UE and its serving STAR-RIS is fixed at $d_c = 75$ m.

\subsection{Validation and Simulations}
\begin{figure*}[t!] 
	\centering
	\subfigure[]{\includegraphics[width=3.1in]{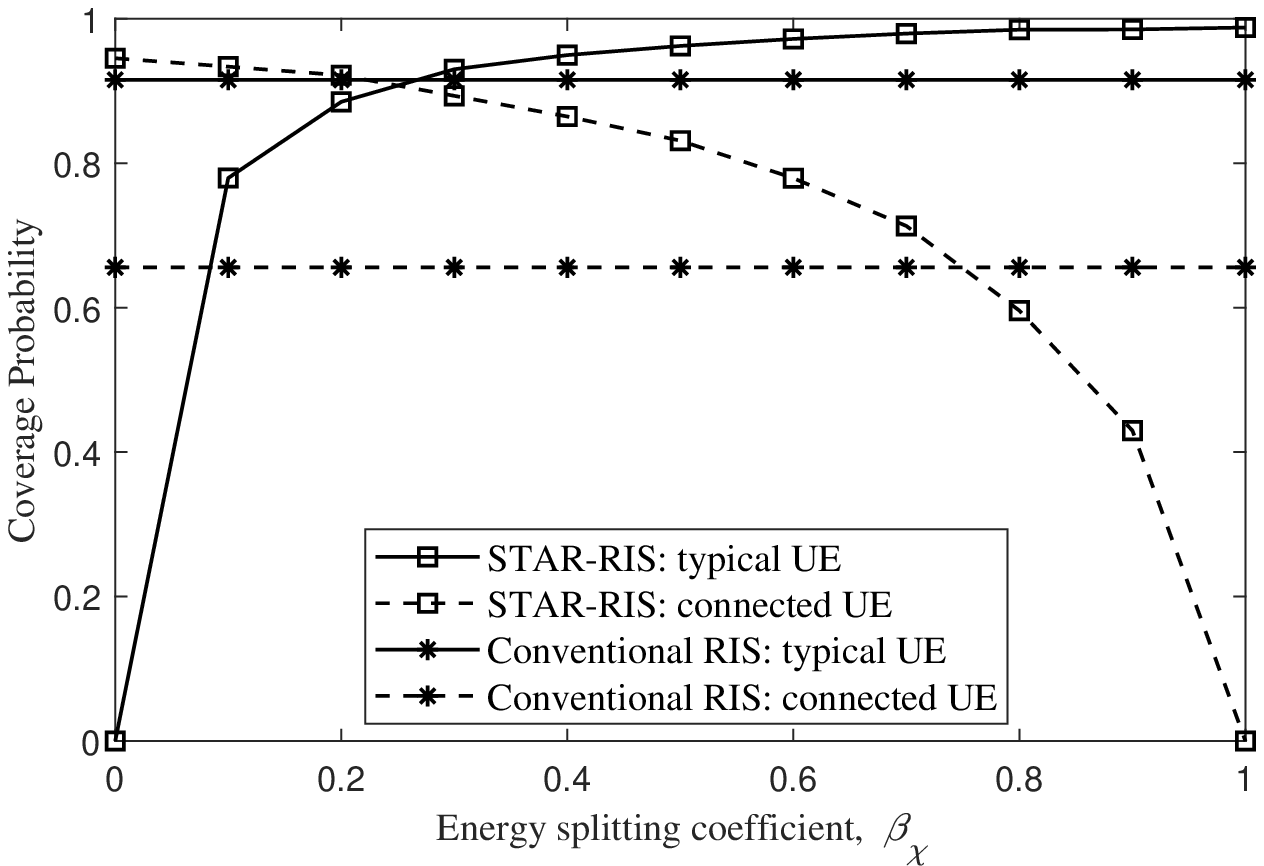}
		\label{e: fig_a}}
	\hfil
	\subfigure[]{\includegraphics[width=3.1in]{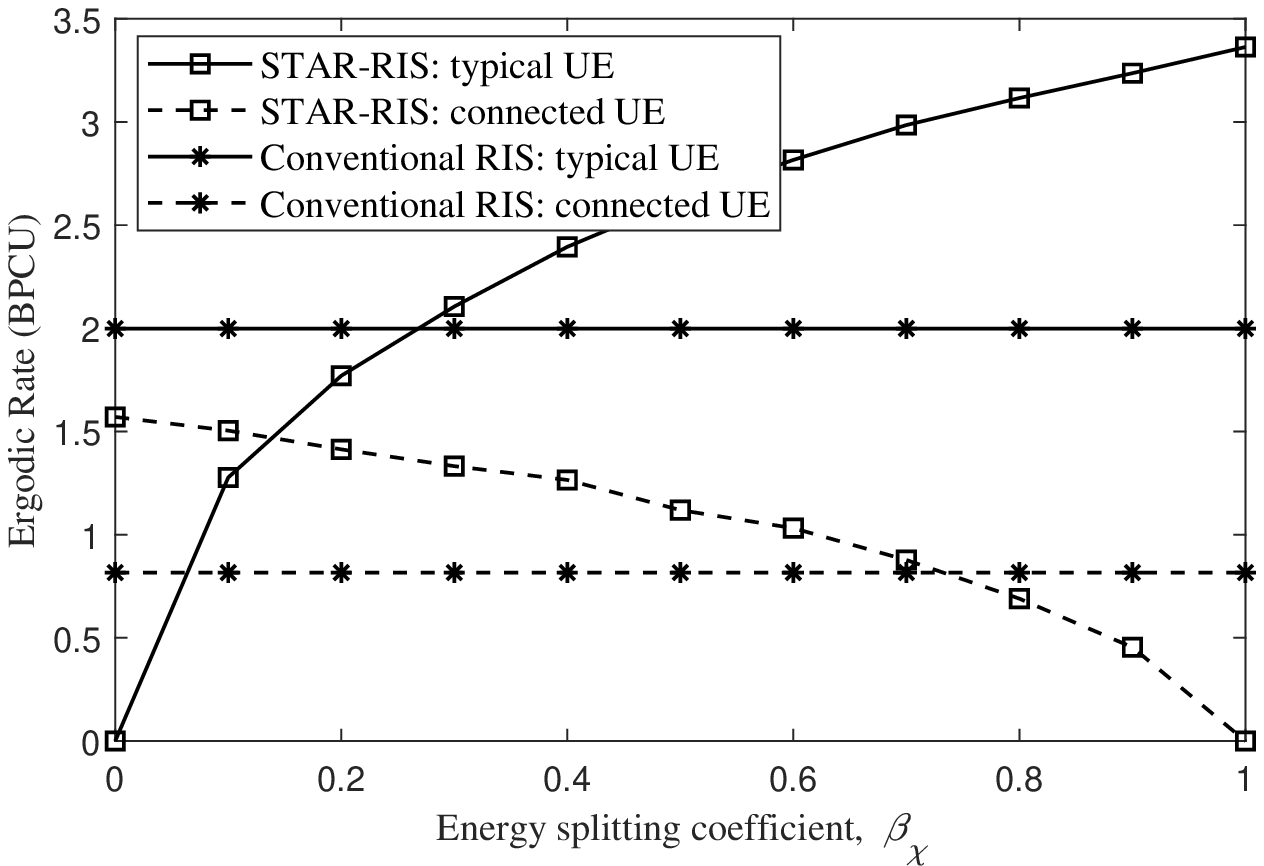}
		\label{e: fig_b}}
	\caption{System performance versus energy splitting coefficient $\beta_\chi$ for the typical UE with $N=16$, $\lambda_B=\lambda_b$, and $\lambda_R=10\lambda_r$: (a) coverage probability; (b) ergodic rate.
		\label{fig: energy splitting UE}}
\end{figure*}
\begin{figure*}[t!] 
	\centering
	\subfigure[]{\includegraphics[width=3.1in]{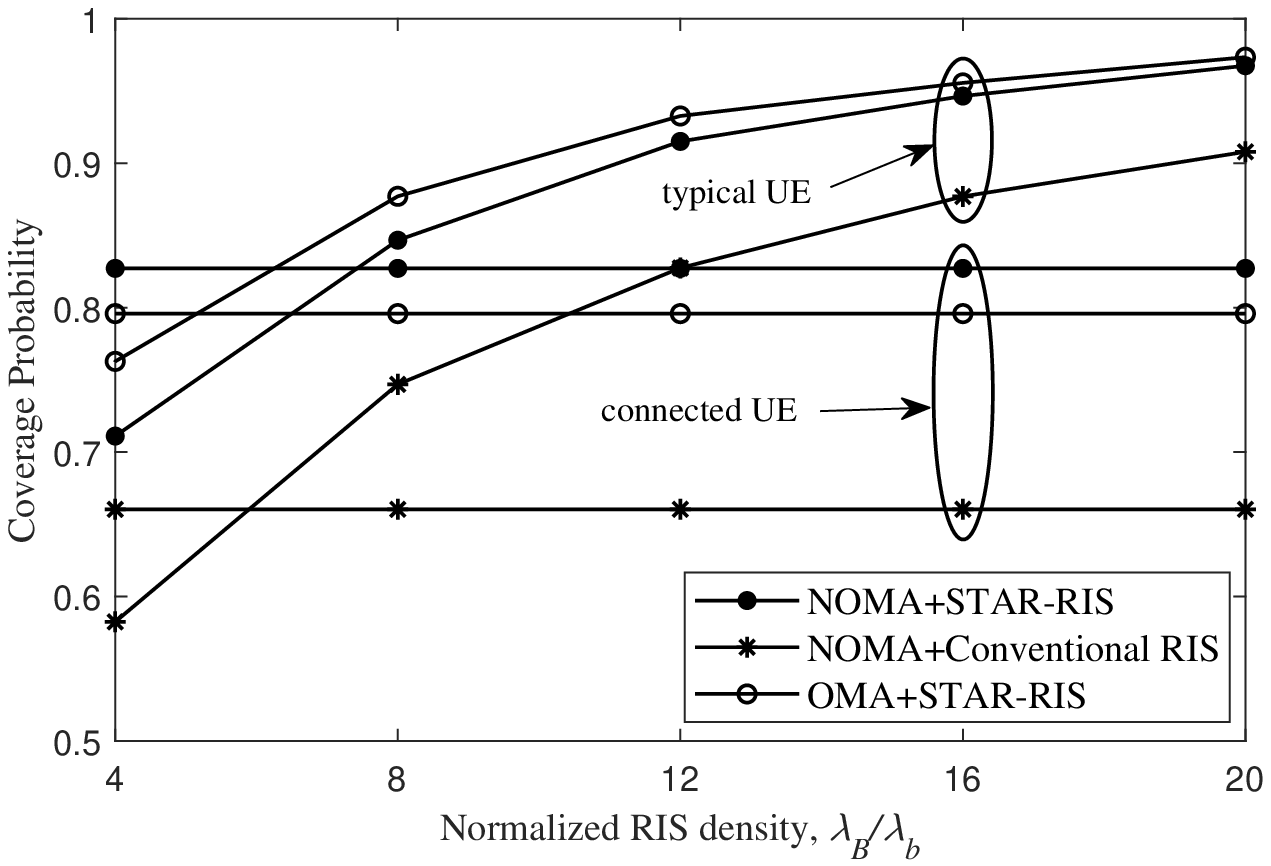}
		\label{d: fig_a}}
	\hfil
	\subfigure[]{\includegraphics[width=3.1in]{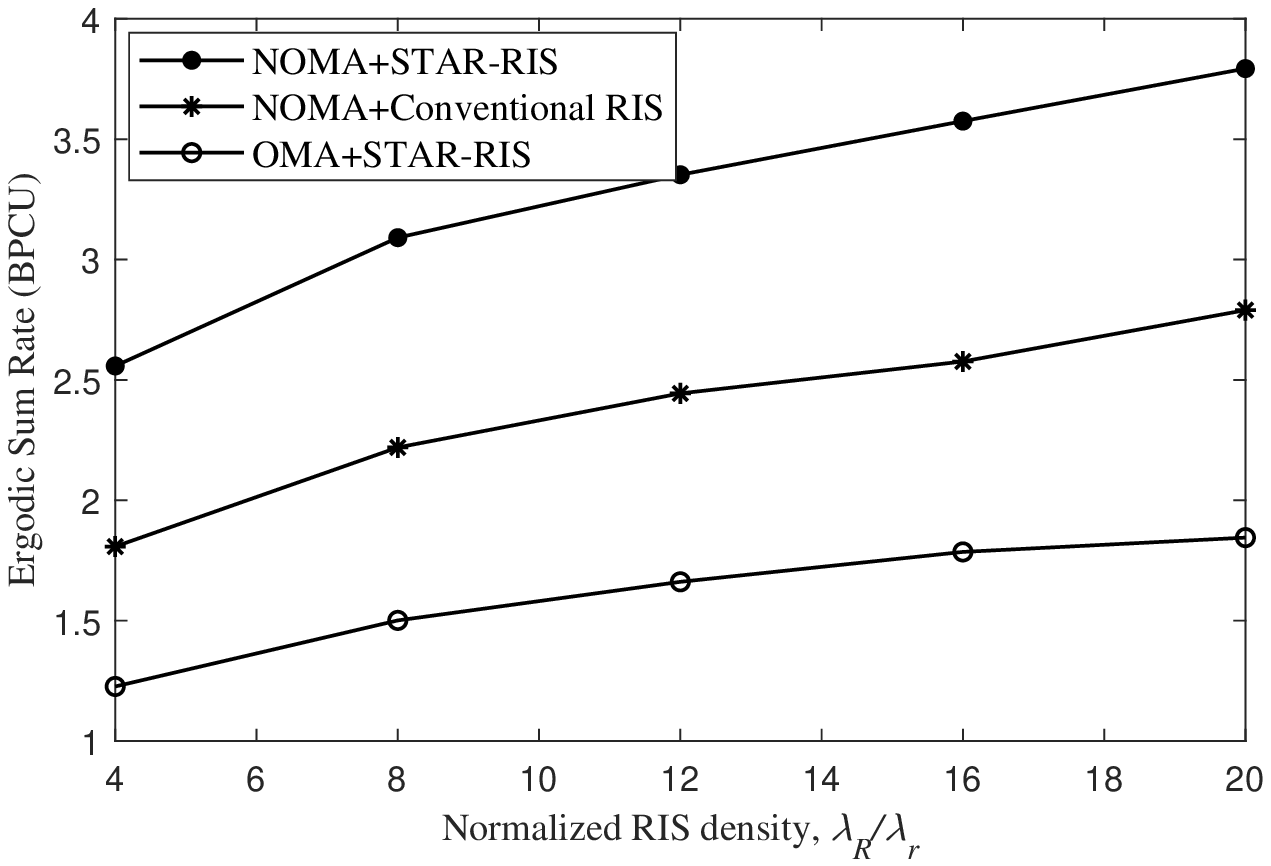}
		\label{d: fig_b}}
	\caption{System performance versus normalized RIS density $\lambda_R/\lambda_r$ with $N=16$, $\lambda_B = \lambda_b$: (a) coverage probability; (b) ergodic sum rate.
	}
	\label{fig: comparison}
\end{figure*}

Validation of analytical expressions of the coverage probability for the typical UE and the connected UE are illustrated in Fig. \ref{fig: verification} with different power allocation coefficients. We use lines to present the analytical results and marks to depict the Monte Carlo simulations. We set the bandwidth $W = 5$ MHz and hence ${n_0}^2 = -107$ dBm. In Fig. \ref{v: fig_a} we vary the SINR threshold for the typical UE while in Fig. \ref{v: fig_b} we vary the counterpart for the connected UE. 
Although the number of the STAR-RIS elements is not large, analytical results fit simulation curves well.  

Due to the limited computing precision of Matlab, it is difficult to calculate the analytical results of the ergodic rate under the predefined parameter setup. However, according to the proof of \textbf{Theorem \ref{theorem: rate typical}} and \textbf{Theorem \ref{theorem: rate connected}} we can find that the expressions of $R_{t,STAR}$ and $R_{c,STAR}$ are from the coverage probabilities of the typical UE and the connected UE, respectively. Therefore, the validation of analytical expressions for the coverage probability guarantees the accuracy of results for the ergodic rate.

\subsection{Impact of Number of Elements}
In this subsection, we investigate the impact of the increase in STAR-RIS elements on both the coverage probability and the ergodic rate. Fig. \ref{fig: element} plots the performance of both the typical UE and the connected UE versus the number of RIS elements $N$. In Fig. \ref{n: fig_a} we focus on the coverage probability while ergodic rate in Fig. \ref{n: fig_b}. We also plot curves when direct BS-UE links are considered in Fig. \ref{n: fig_a} for comparison, where $\lambda_B = \lambda_b$. Fig. \ref{n: fig_a} validates that the impact of the direct links is negligible when the number of elements is large.
As we have discussed in {\bf Remark 2}, the received SINR at NOMA UEs increases with $N$, and hence both the coverage and rate keep growing with the increase of $N$. A design guideline is provided that deploying STAR-RISs with more elements helps to improve the system performance.

\subsection{Impact of Energy Splitting Coefficients}
Here, we focus on the impact of energy splitting coefficients of STAR-RISs. Fig. \ref{fig: energy splitting} plots the coverage probability and the ergodic rate for the paired NOMA UE versus the energy splitting coefficient for transmitting $\beta _{\rm T}$. One can observe that for the STAR-RIS aided scenarios, two kinds of performance are simultaneously maximized when $\beta _{\rm T} = \frac{1}{2}$, which has been discussed in \textbf{Corollary \ref{corollary: opt beta}, \ref{corollary: opt beta c}, \ref{corollary: opt beta rate}} and \textbf{\ref{corollary: opt beta rate c}}. Besides, the curves are symmetric about $\beta _{\rm T} = \frac{1}{2}$.

We denote the energy splitting coefficient of the desired signal for the typical UE as $\beta_\chi$. In Fig. \ref{fig: energy splitting UE}, we plot the coverage probability and the ergodic rate versus $\beta_\chi$. The performance of conventional RIS aided networks is also shown in this figure for comparison. As shown in Fig. \ref{fig: energy splitting UE}, two categories of performance of the paired UEs vary by adjusting the energy splitting coefficient $\beta_\chi$, which means different performance demands can be satisfied. Besides, STAR-RISs with appropriate $\beta_\chi$ help the paired NOMA UEs to achieve better performance than conventional RISs.

\subsection{Comparison among Different Scenarios}
\renewcommand{\theequation}{A.\arabic{equation}}
\setcounter{equation}{0}
\begin{figure*}[ht]
	\normalsize
	\begin{align}\label{A.1}
		{\cal L}_{{I_t}}(s) &= \underbrace{{\mathbb{E}_{\Phi _B}} \! \left[ \! {\prod\limits_{m \in \Phi _B^{\rm T}\backslash i} \!\! {{\mathbb{E}_{|h_t|^2}}\left[ {\exp \left( { - s\beta_{\rm T} {P_B} L_{t,m}^{(k)} |h_t|^2} \right)} \right]} } \right] } _{\mbox {Transmissive Interference}} \nonumber \underbrace{{\mathbb{E}_{\Phi _B}} \! \left[ \! {\prod\limits_{m \in \Phi _B^{\rm R}\backslash i} \!\! {{\mathbb{E}_{|h_t|^2}}\left[ {\exp \left( { - s\beta_{\rm R} {P_B} L_{t,m}^{(k)} |h_t|^2} \right)} \right]} } \right] } _{\mbox {Reflective Interference}} \nonumber \\
		&\overset{(a)}{=} \prod\limits_{\chi \in \{{\rm T},{\rm R}\}} \exp \left( { - \pi \lambda _B\int_{r_t}^\infty  {\left( {1 - {\mathbb{E} _{|{\tilde h}_t|^2}}\left[ {\exp \left( { - s\theta_I{C_r}\beta_{\chi}{P_B} \left(rd_t \right)^{-\alpha_r} 
							|{\tilde h}_t|^2} \right)} \right]} \right)rdr} } \right) \nonumber \\
		&\overset{(b)}{=} \prod\limits_{\chi \in \{{\rm T},{\rm R}\}} \exp \left( { - \pi \lambda _B \int_{r_t}^\infty  {\left( {1 - \left( 1 + s\theta_I{C_r}\beta_{\chi}{P_B} \left(rd_t \right)^{-\alpha_r}  \right)^{- 1}} \right)rdr} } \right) ,
	\end{align}
	\hrulefill \vspace*{0pt}
\end{figure*}
In Fig. \ref{d: fig_a}, we plot coverage probabilities versus normalized RIS density $\lambda_R/\lambda_r$ in multiple scenarios, where different kinds of RISs and multiple access techniques are considered. We observe that the STAR-RIS outperforms the conventional RIS with the increase of $\lambda_R/\lambda_r$ because the stronger desired signal can be transmitted or reflected by the STAR-RIS. Compared with OMA, NOMA enhances the coverage probability of the connected UE. This enhancement comes from the higher power allocated to the connected UE which has a high probability to be in a worse channel condition than the typical UE in NOMA systems.

In Fig. \ref{d: fig_b}, we compare ergodic sum rates of the paired UEs versus normalized RIS density $\lambda_R/\lambda_r$ for STAR-RIS aided NOMA networks, conventional RIS aided NOMA networks, and STAR-RIS aided OMA networks. We can observe that the NOMA system always outperforms the OMA system because of its high bandwidth efficiency. Similar to the observation in coverage probability, STAR-RISs with the optimal energy splitting coefficients achieve a higher ergodic sum rate than conventional RISs. Therefore, with appropriate energy splitting coefficients, STAR-RISs have the best performance among all scenarios we considered in this subsection.

\section{Conclusion}
In this paper, a fitting method has been proposed to approximate the distribution of the RIS aided composite CPG. Then, a general analytical framework has been provided to evaluate the coverage probability and the ergodic rate of STAR-RIS aided NOMA multi-cell networks. Theoretical expressions in conventional RIS-aided networks have been obtained for comparison. For more insights, we have investigated the impact of energy splitting coefficients and considered the interference-limited scenario as a special case. The analytical results have revealed that appropriate energy splitting coefficients can simultaneously improve the system coverage and the ergodic performance. The numerical results have shown that: 1) the increase of RIS elements helps to improve the system coverage and the rate performance; 2) a specific range of energy splitting coefficients guarantees STAR-RISs outperform conventional RISs; 3) STAR-RISs provide flexibility for satisfying different UE demands by altering energy splitting coefficients.

\numberwithin{equation}{section}
\section*{Appendix~A: Proof of Lemma~\ref{lemma: Lap I_r}}
\label{Appendix:A}
\renewcommand{\theequation}{A.\arabic{equation}}
\setcounter{equation}{1}
Based on Campbell’s theorem, the Laplace transform of the interference for the typical UE can be expressed as \eqref{A.1}, where $|{\tilde h}_t|^2 \sim \Gamma({1,1})$. $(a)$ is obtained by using the probability generating functional (PGFL) and the fact that $\Gamma({k,\theta}) = k\theta \Gamma({k,\frac{1}{k}})$. $(b)$ follows from the moment generation function of the Gamma distribution. By applying \cite[eq. C.2]{1multicell2} in our previous work \cite{1multicell2}, we can obtain a more elegant form as shown in \eqref{eq: Lap I_r}. 

\numberwithin{equation}{section}
\section*{Appendix~B: Proof of Theorem~\ref{theorem: cp typical}}
\label{Appendix:B}
\renewcommand{\theequation}{B.\arabic{equation}}
\setcounter{equation}{0}

We denote $Z = \tau_t^* \frac{I_t + {n_0}^2}{\theta_r \beta_\chi P_B L_{t,i}^{(k)}}$. Based on the fact that $\frac{1}{\theta_r} |h_r|^2 \overset{appr.}{\sim} \Gamma({\bar k}_r,1)$ as well as the CDF of the Gamma random variable, we have
\begin{align}
	P_{t,STAR,\chi}|_{d_{t,i}^{(k)}} &= \mathbb{E} _{I_t} \left[ e^{-Z} \sum_{m=0}^{{\bar k}_r-1} \frac{Z^m}{m!} \right] \nonumber \\ &= \sum_{m=0}^{{\bar k}_r-1}\frac{(-1)^m}{m!} \left[ \frac{\partial ^m}{ \partial x^m} \mathbb{E} _{I_t} \left[e^{-Zx} \right] \right]_{x=1}.
\end{align}

When the signal transmission mode of assisted STAR-RIS for the typical UE is $\chi$, the coverage probability is expressed as
\begin{align}\label{eq: cp 1}
	&P_{t,STAR,\chi} = \int_{0}^\infty \int_{0}^\infty  f_{BR}(r_1)f_{RU}(r_2) \sum_{m=0}^{{\bar k}_r-1}\frac{(-1)^m}{m!}  \nonumber \\ & \times \left[ \frac{\partial ^m}{ \partial x^m} {\cal L}_{I_r}(s_{t,\chi} x)\exp(-s_{t,\chi} {n_0}^2x) \right]_{x=1}  dr_1dr_2,
\end{align}
where $s_{t,\chi} = \frac{\tau_t^*(r_1r_2)^{\alpha_r}}{\theta_r \beta_\chi P_B C_r}$.

For simplicity, We denote $D_{{\rm T}\chi} = \frac{\theta_I\tau_t^*\beta _{\rm T}}{\theta_r\beta _\chi}$ and $D_{{\rm R}\chi} = \frac{\theta_I\tau_t^*\beta _{\rm R}}{\theta_r\beta _\chi}$. Then we plug \eqref{eq: BR pdf} and \eqref{eq: Lap I_r} into \eqref{eq: cp 1}. After some simple algebraic manipulations, the coverage probability is derived as
\begin{align}
	P_{t,STAR,\chi} &= \int_{0}^\infty \int_{0}^\infty {2\pi \lambda_B r_1 f_{RU}(r_2)}  \nonumber \\ & \times \sum_{m=0}^{{\bar k}_r-1}\frac{(-1)^m}{m!} \left[ \frac{\partial ^m}{ \partial x^m} \exp \left(V_{\chi}(x)\right) \right]_{x=1} dr_1 dr_2,
\end{align}
where
\begin{align}\label{eq: Vx}
	V_{\chi}(x) = &-s_{t,\chi}{n_0}^2x - \frac{1}{2}\pi \lambda_B {r_1}^2 \xi_0 \left( 1,\alpha_r, D_{{\rm T}\chi}; x \right) \nonumber \\ & - \frac{1}{2}\pi \lambda_B {r_1}^2 \xi_0 \left( 1,\alpha_r, D_{{\rm R}\chi}; x \right).
\end{align}

Now let us focus on the high-order derivatives of the composite function $\exp \left( V_{\chi}(x) \right)$. According to Fa\`a di Bruno's formula \cite{2Faa}, we have
\begin{align}
	&\frac{\partial ^m}{\partial x^m} \exp \left( V_{\chi}(x) \right) \nonumber \\ &  = \sum_{l=0}^{m} \exp \left( V_{\chi}(x) \right)^{(l)} {\cal B}_{m,l} \left( V_{\chi}^{(1)}(x), ..., V_{\chi}^{(m-l+1)}(x) \right) \nonumber \\
	& \overset{(a)}{=} \exp \left( V_{\chi}(x) \right) {\cal B}_{m} \left( V_{\chi}^{(1)}(x), ..., V_{\chi}^{(m)}(x) \right),
\end{align}
where ${\cal B}_m \left(x_1, ..., x_m \right)$ is the $m$th complete Bell polynomial. ${\cal B}_{m,l} \left(x_1, ..., x_{m-l+1} \right)$ is the incomplete Bell polynomial. $(a)$ is obtained by combining the fact that $\exp(x)^{(m)} = \exp(x)$ and the property that ${\cal B}_m \left(x_1, ..., x_m \right) = \sum_{l=1}^{m} {\cal B}_{m,l} \left(x_1, ..., x_{m-l+1} \right)$.

By applying \cite[eq. 15.5.2]{2Nist}, the $m$th derivative of $V_{\chi}(x)$ has the following closed form 
\begin{align}
	V_{\chi}^{(m)}(x) =& -\Delta^{(m)}  - \frac{1}{2}\pi \lambda_B {r_1}^2 \xi_m \left(1,\alpha_r, D_{\chi}^{\rm T}; x \right) \nonumber \\ & - \frac{1}{2}\pi \lambda_B {r_1}^2 \xi_m \left( 1,\alpha_r, D_{\chi}^{\rm R}; x \right) ,
\end{align}
where $\Delta^{(0)} = s_{t,\chi}{n_0}^2x$, $\Delta^{(1)} = s_{t,\chi}{n_0}^2$, and $\Delta^{(m)} = 0$ when $m \ge 2$.

According to the location relationship among the typical UE, the assisted STAR-RIS, and 
the serving BS, if the STAR-RIS aided link is transmissive, the BS can be only located in the half part of the considered plane $\mathbb{R}^2$ split by the STAR-RIS and vice versa. Therefore, the overall coverage probability of the typical UE is expressed as
\begin{align}
	P_{t,STAR} &= \frac{1}{2}P_{t,STAR,{\rm T}} + \frac{1}{2}P_{t,STAR,{\rm R}}.
\end{align}

Then, the proof is completed.

\numberwithin{equation}{section}
\section*{Appendix~C: Proof of Corollary~\ref{corollary: opt beta}}
\label{Appendix:C}
\renewcommand{\theequation}{C.\arabic{equation}}
\setcounter{equation}{0}

Let us denote $\beta = \beta _{\rm T} = 1- \beta_{\rm R}$. Since $z$ makes no difference here, we denote $V_{\chi}(x)\triangleq V_{\chi}(z,x)$. We denote $c_0 = \frac{ \tau_t^*(r_1r_2)^{\alpha_r}{n_0}^2}{\theta_r  P_B C_r}$. By substituting $s_\chi = \frac{\tau_t^*(r_1r_2)^{\alpha_r}}{\theta_r \beta_\chi P_B C_r}$ into \eqref{eq: Vx}, 
$V_{\chi}(x)$ is rewritten as
\begin{align}\label{eq: lap 1}
	V_{\chi}(x;\beta) = & -\frac{c_0x}{\beta_{\chi}} - \frac{1}{2}\pi {r_1}^2\lambda_B  \xi_0 \left( 1,\alpha_r, \frac{\theta_I\tau_t^*xz}{\theta_r}; \frac{\beta }{\beta_{\chi}} \right) \nonumber \\ & - \frac{1}{2}\pi {r_1}^2\lambda_B\xi_0 \left( 1,\alpha_r, \frac{\theta_I\tau_t^*xz}{\theta_r}; \frac{(1-\beta) }{\beta_{\chi}} \right).
\end{align}

To find the optimal energy splitting coefficient for the coverage probability, we take the derivative of $P_{t,STAR}$ with respect to $\beta$
\begin{align}\label{eq: C.2}
	&\frac{\partial}{\partial \beta}P_{t,STAR} \nonumber \\ & = \frac{\partial}{\partial \beta} \int_{0}^\infty \int_{0}^\infty \! \pi \lambda_B r_1 f_{RU}(r_2) \nonumber \\ & \times \sum_{m=0}^{k_r-1}\frac{(-1)^m}{m!} \left[ \frac{\partial ^m}{ \partial x^m} \exp \left(V_{\rm T}(x;\beta) + V_{\rm R}(x;\beta) \right) \right]_{x=1} \!\!\!\!\!\! dr_1 dr_2 \nonumber \\
	& \overset{(a)}{=} \int_{0}^\infty \int_{0}^\infty \pi \lambda_B r_1 f_{RU}(r_2) \nonumber \\ & \times \!\! \sum_{m=0}^{k_r-1}\frac{(-1)^m}{m!} \left[ \frac{\partial ^m}{ \partial x^m} \frac{\partial}{\partial \beta} \exp \left(V_{\rm T}(x;\beta) + V_{\rm R}(x;\beta) \right) \right]_{x=1}\!\!\!\!\!\!\!\!\! dr_1 dr_2 .
\end{align}
where $(a)$ utilizes the fact that the high-order derivatives of $\exp \left( V_{\chi}(x;\beta) \right)$ are continuous, and hence the operators of different partial derivatives are interchangeable. 

Now let us calculate the first derivative in \eqref{eq: C.2}, we have
\begin{align}
	&\frac{\partial}{\partial \beta} \exp \left(V_{\rm T}(x;\beta) + V_{\rm R}(x;\beta) \right) \nonumber \\ & \triangleq \frac{\partial}{\partial \beta} \exp \left(V(x;\beta) \right) = \exp \left(V(x;\beta) \right) {\cal B}_1 \left( V^{(1)}(x;\beta) \right) \nonumber \\
	&= \exp \left(V(x;\beta) \right) \left( -c_0x \left( \frac{1}{(1-\beta)^2} -\frac{1}{\beta^2} \right) \right. \nonumber \\ &+ \left. \frac{ \pi {r_1}^2\lambda_B}{\beta^2}  \xi_1 \left( k_r,\alpha_r, \tau_t^*x; \frac{1-\beta }{\beta} \right) \right. \nonumber \\ & - \left. \frac{ \pi {r_1}^2\lambda_B}{(1-\beta)^2}  \xi_1 \left( k_r,\alpha_r, \tau_t^*x; \frac{\beta }{1-\beta} \right) \right).
\end{align}

It can be observed that when $\beta = \frac{1}{2}$, $\frac{\partial}{\partial \beta}P_{t,STAR} = 0$. Moreover, when  $\beta \to\frac{1}{2}^-$, $\frac{\partial}{\partial \beta}P_{t,STAR} > 0$, and when  $\beta \to\frac{1}{2}^+$, $\frac{\partial}{\partial \beta}P_{t,STAR} < 0$. Thus, $P_{t,STAR}$ is maximized when $\beta_{\rm T} = \beta_{\rm R} = \frac{1}{2}$.

\numberwithin{equation}{section}
\section*{Appendix~D: Proof of Theorem~\ref{theorem: rate typical}}
\label{Appendix:D}
\renewcommand{\theequation}{D.\arabic{equation}}
\setcounter{equation}{0}

If $\gamma_{t \to c} > \tau_c$ holds, we have $\gamma_t > a_t \tau_c^*$. Therefore, \eqref{eq: rate typical def} can be rewritten as $R_{t,STAR}= \int_{z > a_t \tau_c^*} \log_2 \left( 1  + \gamma_t \right) \frac{\partial }{\partial z}\left(1 - \mathbb{P} \left(\gamma_t > z \right) \right) dz$.
%\begin{align}\label{eq: D1}
%	R_{t,STAR}= \int_{z > a_t \tau_c^*} \log_2 \left( 1  + \gamma_t \right) \frac{\partial }{\partial z}\left(1 - \mathbb{P} \left(\gamma_t > z \right) \right) dz	.
%\end{align}
We denote ${\bar F}_t(z) = \mathbb{P} \left(\gamma_t > z \right)$, which represents the CCDF of the decoding SINR for the typical UE. Then we can calculate
\begin{align}
	R_{t,STAR} &= - \int_{a_t \tau_c^*}^\infty \int_0^z \frac{1}{\ln 2 (1+x)} dx  \frac{\partial }{\partial z}{\bar F}_{t,STAR}(z) dz	\nonumber \\
	&\overset{(a)}{=} -\frac{1}{\ln 2} \int_0^{a_t \tau_c^*} \frac{1}{1+x} \int_{a_t \tau_c^*}^\infty \frac{\partial }{\partial z}{\bar F}_{t,STAR}(z) dz dx  \nonumber \\ &-\frac{1}{\ln 2} \int_{a_t \tau_c^*}^\infty \frac{1}{1+x} \int_x^\infty \frac{\partial }{\partial z}{\bar F}_{t,STAR}(z) dz dx	\nonumber \\
	&\overset{(b)}{=} \log_2 \left( 1  + a_t \tau_c^* \right) {\bar F}_{t,STAR}(a_t \tau_c^*) \nonumber \\ & + \frac{1}{\ln 2} \int_{a_t \tau_c^*}^\infty \frac{{\bar F}_{t,STAR}(x)}{1+x} dx,
\end{align}
where $(a)$ is obtained by exchanging the order of integration. $(b)$ is obtained by using the fact that $\lim \limits_{z \to \infty } {\bar F}_{t,STAR}(z) = 0$.

Now we calculate ${\bar F}_{t,STAR}(z)$. Based on \eqref{eq: SINR typical}, we have
\begin{align}
	&{\bar F}_{t,STAR}(z) = \frac{1}{2} \nonumber \\ & \times \!\!\!\! \sum_{\chi \in \{{\rm T,R}\}} \int_{0}^\infty \!\! \int_{0}^\infty \mathbb{P} \left( |h_t|^2 >  \frac{ \left( I_t + {n_0}^2 \right)z }{a_t\beta_\chi P_BC_r (r_1r_2)^{-\alpha_r} }   \right) d{r_1} d{r_2}.
\end{align}

By utilizing the same proof as Theorem \ref{theorem: cp typical}, we can obtain \eqref{eq: rate typical}. This proof is completed.

%\appendices

%\section{}

%\section*{Acknowledgment}


%\ifCLASSOPTIONcaptionsoff
 % \newpage
%\fi

\bibliographystyle{IEEEtran}

\bibliography{reference}


\end{document}